\documentclass[prl,twocolumn,showpacs,superscriptaddress,preprintnumbers,amssymb]{revtex4}
\usepackage{graphicx}
\usepackage{dcolumn}
\usepackage{bm}
\usepackage{wasysym}

\newcommand{\beq}{\begin{equation}}
\newcommand{\eeq}{\end{equation}}
\newcommand{\beqn}{\begin{eqnarray}}
\newcommand{\eeqn}{\end{eqnarray}}

\begin{document}
\title{Classical Phase Transitions of Geometrically Constrained O($N$) Spin Systems}
\author{Cenke Xu}
\affiliation{Department of Physics, Harvard University, Cambridge,
MA 02138}
\date{\today}

\begin{abstract}

We study the phase transition between the high temperature
algebraic liquid phase and the low temperature ordered phase in
several different types of locally constrained O($N$) spin
systems, using a unified constrained Ginzburg-Landau formalism.
The models we will study include: $\mathbf{1}$, O($N$) spin-ice
model with cubic symmetry; $\mathbf{2}$, O($N$) spin-ice model
with easy-plane and easy-axis anisotropy; $\mathbf{3}$, a novel
O($N$) ``spin-plaquette" model, with a very different local
constraint from the spin-ice. We calculate the renormalization
group equations and critical exponents using a systematic
$\epsilon = 4 - d$ expansion with constant $N$, stable fixed
points are found for large enough $N$. In the end we will also
study the situation with softened constraints, the defects of the
constraints will destroy the algebraic phase and play an important
role at all the transitions.

\end{abstract}
 \maketitle

\section{I, introduction}

It is well-known that sometimes constraints imposed locally can
lead to stable phases with unusual long distance correlations. For
instance, at zero temperature the local constraints can give rise
to quantum Bose liquid phase with gapless excitations analogous to
photon or even graviton
\cite{sondhiphoton,hermelephoton,wenphoton,xugraviton1,xugraviton2}.
At finite temperature, in three dimensional classical dimer model
(CDM) and spin-ice systems, the local geometric constraint grants
the high temperature spin disordered phase an algebraic power-law
correlation between dimer and spin densities
\cite{sondhidipole,sondhispinice2,sondhidimer,balentsdimer,dimersimulation1,dimersimulation2},
instead of the standard short-range correlations according to the
Ginzburg-Landau theory. This algebraic phase is usually called the
Coulomb phase. In both the dimer models and spin-ices, physical
quantities are defined on links of a bipartite lattice, and let us
take the cubic lattice for simplicity. The ensemble of 3d CDM is
all the configurations of dimer coverings, which are subject to a
local constraint on every site: each site is connected to
precisely $m$ dimers with $0 < m < 6$ (denoted as CDM-$m$), and
most studies have been focused on the case with $m = 1$. The
ensemble of the spin-ice model is all the configurations of spins
on the links, and the sum of the six spin vectors around each site
is zero, which is usually called the ice-rule \cite{pauling}.
Although the spin is generically an O(3) vector, due to the
spin-orbit coupling, the spins in the spin-ice materials prefer to
align along the links where the spins reside, therefore people
usually treat the spins in spin-ice materials Ising spins
\cite{sondhispinice2}, hence the spin-ice model on the cubic
lattice is mathematically equivalent to CDM-3. The power-law
correlation of the Coulomb phase has been confirmed by numerical
simulations \cite{balentsdimer,dimersimulation1,dimersimulation2}
and also neutron scattering in spin-ice materials such as
$\mathrm{Ho_2Ti_2O_7}$ and $\mathrm{Dy_2Ti_2O_7}$
\cite{spinicemonopole1}.

The partition function of the CDM-$m$ is a sum of all the dimer
configurations allowed by the constraint, with a Boltzmann weight
that favors certain dimer configurations. To describe CDM-$m$
concisely, we can introduce the ``magnetic field" $B_{i, \mu} =
(n_{i, \mu} - m/6)\eta_i$ with $\mu = +\hat{x}, +\hat{y},
+\hat{z}$, where $\eta_i = (-1)^i$ is a staggered sign
distribution on the cubic lattice. The number $n_{i, \mu}$ is
defined on each link $(i, \mu)$ between sites $i$ and $i + \mu$,
and $n_{i,\mu} = 1, \
0$ represents the presence and absence of dimer. 
Now the local constraint of the dimer system can be rewritten as a
Gauss law constraint $\vec{\nabla}\cdot \vec{B} = 0$. The standard
way to solve this Gauss law constraint is to introduce vector
potential $\vec{A}$ defined on the unit plaquettes of the cubic
lattice, and $\vec{B} = \vec{\nabla}\times \vec{A}$
\cite{sondhi2003,Hermele2005}. Since vector $\vec{A}$ is no longer
subject to any constraint, it is usually assumed that at low
energy the system can be described by a local field theory of
$\vec{A}$, for instance the low energy field theory of the Coulomb
phase reads \beqn F \sim \int d^3x (\vec{\nabla}\times \vec{A})^2
+ \cdots \label{coulombfield}\eeqn which is invariant under gauge
transformation $\vec{A} \rightarrow \vec{A} + \vec{\nabla}f$, $f$
is an arbitrary function of space. In the Coulomb phase, the
correlation of magnetic field $B_\mu$ in the momentum space reads
\beqn \langle B_{\mu}(\vec{q}) B_{\nu}(-\vec{q}) \rangle  \sim
\delta_{\mu\nu} - \frac{q_\mu q_\nu}{q^2}.\eeqn

The spin-ice model can be straightforwardly generalized to the
O($N$) case. We can define an O($N$) spin vector $S^a$ with unit
length $\sum_a (S^a)^2 = 1$ on each link $(i,\mu)$ of the cubic
lattice (Fig. \ref{dimeron}$a$), with $a = 1 \cdots N$, and we
assume that the largest term of the Hamiltonian imposes an
ice-rule constraint \cite{pauling} for O($N$) spins on the six
links shared by every site: \beqn \sum_{\mu = x, y, z} S^a_{i,\mu}
+ S^a_{i - \mu, \mu}= 0. \label{onconstraint}\eeqn The magnetic
field formalism developed for the CDM can be naturally applied
here, with $\phi^a_{i, \mu} = \eta_i S^a_{i, +\mu}$, and the
constraint Eq. \ref{onconstraint} can be written as \beqn \sum_\mu
\nabla_\mu \phi^a_\mu = 0. \label{phiconstraint} \eeqn Then in the
Coulomb phase the momentum space correlation between $\phi^a_\mu$
reads \beqn \langle \phi^a_\mu(\vec{q})\phi^b_\nu (-\vec{q})
\rangle \sim \delta_{ab}(\delta_{\mu\nu} - \frac{q_\mu
q_\nu}{q^2}), \eeqn which after Fourier transformation leads to
the same $1/r^3$ power-law correlation as the dimer model.

Another constrained spin system, which is very different from the
spin-ice is the classical plaquette model (CPM). In Ref.
\cite{pankov2007,xuwu2008} a plaquette model engineered from the
SU(4) Heisenberg model on the cubic lattice was studied, and one
of the twelve unit square faces shared by each site of the cubic
lattice is occupied by a plaquette, which is physically the SU(4)
singlet formed by four fermions at the corners. The exact SU(4)
symmetry can be realized in cold atom system without fine-tuning
\cite{wu2003,gorshkov2009}. The ensemble of the CPM-$m$ is all the
plaquette configurations on the cubic lattice, with exactly $m$
plaquettes shared by each site (Fig. \ref{dimeron}$b$). To
generalize this system to O($N$), let us define O($N$) spins $S^a$
on the unit faces instead of the links of the cubic lattice, and
impose the constraint that the sum of spin vectors on all the
twelve faces shared by every site be zero (Fig. \ref{dimeron}$c$).
When $N = 1$, this model is equivalent to CPM-6. In this case it
is most convenient to introduce symmetric rank-2 tensor
$\phi^a_{i, \mu\nu} = \eta_i S^a_{i, \mu\nu}$ with $\mu, \nu = +
\hat{x}, + \hat{y}, + \hat{z}$. $(i,\mu\nu)$ denotes the unit
square face shared by sites $i$, $i + \mu$, $i + \nu$ and $i + \mu
+ \nu$. In terms of $\phi^a_{\mu\nu}$, the spin plaquette
constraint can be written as \beqn \sum_{\mu \neq \nu}
\nabla_\mu\nabla_\nu \phi^a_{\mu\nu} = 0.
\label{phi2constraint}\eeqn This spin-plaquette system also has an
algebraic phase at finite temperature, with the following momentum
space correlation: \beqn \langle
\phi^a_{\mu\nu}(\vec{q})\phi^b_{\rho\sigma}(-\vec{q}) \rangle
&\sim& \delta_{ab}(\delta_{\mu\rho}\delta_{\nu\sigma} +
\delta_{\mu\sigma}\delta_{\nu\rho} \cr\cr &-& \frac{q_\mu q_\nu
q_\rho q_\sigma}{q_x^2q_y^2 + q_y^2 q_z^2 + q_x^2q_z^2}). \eeqn

Besides the high temperature algebraic phases in the models
discussed above, at low temperature spins are expected to order
according to the details of the Hamiltonian. In this work we will
study the phase transition between the high temperature algebraic
phase and the low temperature spin ordered phase in O($N$)
spin-ice (section II), with both cubic symmetry and anisotropies,
as well as O($N$) spin-plaquette system (section III). In section
IV we will study the situation with softened constraints. The
nature of the transitions clearly depend on the low temperature
spin order pattern, and in this work we will focus on one
particular type of spin order, which has nonzero net $\langle
\phi^a_\mu \rangle$ and $\langle \phi^a_{\mu\nu} \rangle$ at large
length scale. Our calculation will be based on $\epsilon = 4 - d$
expansion. Due to the complexity of the calculation, in our
current work we will keep the precision to the first order
$\epsilon$ expansion.

\section{II, O($N$) spin-ice model}

\subsection{A, Brief Review of the CDM}

Let us first give a brief review of the previously studied phase
transition of the CDM, or the spin-ice model with $N = 1$. The
field theory Eq. \ref{coulombfield} misses one important piece of
information: the magnetic field $\vec{B}$ and the vector potential
$\vec{A}$ are both discrete. Therefore mathematically we should
introduce ``vertex operator" $L_v \sim \sum_\mu \cos[2\pi( A_\mu +
\tilde{a}_\mu)]$ to the field theory Eq. \ref{coulombfield}, and
$\tilde{a}_\mu$ is a nonzero background distribution of the vector
potential, which is introduced for any nonzero $m$. The Coulomb
phase is a phase where this vertex operator is irrelevant
perturbatively. The vertex operator will become nonperturbative
and drive a phase transition when it is large, and in order to
describe this phase transition one can introduce matter fields in
the vertex operator which couple to the gauge field minimally:
\beqn L_v \sim \sum_\mu \cos[2\pi (\nabla_\mu \theta - A_\mu -
\tilde{a}_\mu)], \label{vertexmatter}\eeqn $\theta$ is the phase
angle of the matter field $\psi \sim e^{i\theta}$. Due to the
existence of the nonzero $\tilde{a}_\mu$, the matter fields $\psi$
move on a nonzero background magnetic field, the band structure of
the matter fields have multiple minima in the Brillouin zone, and
the transformation between these minima encodes the information of
the lattice symmetry \cite{senthilmotrunich}. Therefore in
addition to manifesting the discrete nature of the gauge potential
$A_\mu$, the condensation of the matter field leads to lattice
symmetry breaking, which corresponds to the crystal phase of the
CDM. For instance the transition between the Coulomb and columnar
crystal phases of the CDM-1 model is described by the CP(1) model
with an enlarged SU(2) global symmetry
\cite{powell1,balentsdimer}. This field theory is highly
unconventional, in the sense that it is not formulated in terms of
physical order parameters. It is expected that more general
CDM-$m$ models can also be described by similar Higgs transition,
although the detailed lattice symmetry transformation for matter
fields would depend on $m$.

One might be tempted to describe the transition between the
Coulomb and columnar phases of CDM-1 trough a Ginzburg-Landau
approach. One can introduce an O(3) vector $\vec{\varphi}$ with
cubic symmetry anisotropy in favor of six axial directions, and
$\langle \varphi_\mu \rangle \sim \pm 1$ represents six fold
degenerate columnar order. However, the hedgehog monopole
configuration of the O(3) vector $\vec{\varphi}$ always involves a
broken dimer $i.e.$ a defect of the constraint. Then as long as we
forbid the presence of the defects, this O(3) model is
monopole-free, and it is well-known that the monopole-free O(3)
nonlinear sigma model is equivalent to the CP(1) model
\cite{motrunichashvin}, which has very different critical
exponents from the O(3) Wilson-Fisher fixed point
\cite{murthy1993,motrunichashvin2}.

\subsection{B, Isotropic O($N$) Spin-ice}

When $N > 1$, the spin takes continuous values, therefore the
formalism of the phase transition based on vertex operator in the
previous section is no longer applicable. Also, it is impossible
to write down a vertex operator with the O($N$) spin symmetry.
Therefore, we need to seek for a different formalism. We consider
the following Hamiltonian for O($N$) spin-ice in addition to the
dominant constraint Eq. \ref{onconstraint}: \beqn E &=&
\sum_{i,\mu, a} J_1 S^a_{i -\mu, \mu } S^a_{i,\mu} +
\sum_{i,a}\sum_{\mu\neq\nu} J_2 S^a_{i,\mu}S^a_{i + \nu,\mu}.
\label{model}\eeqn $J_1$ is a Heisenberg coupling between spins
along the same lattice axis, $J_2$ is a Heisenberg coupling
between spins on two parallel links across a unit square face. If
$J_1 > 0$ and $J_2 < 0$, in the ground state spins are
antiparallel along the same axis, but parallel between parallel
links across a unit square $i.e.$ $\langle S^a_{i + \mu} \rangle
\sim (-1)^{i_\mu} \mathcal{S}^a_\mu$, $\mathcal{S}^a_\mu$ is a
constant O($N$) vector. Using the CDM terminology, we will call
this state the columnar state. If $J_1$ and $J_2$ are both
positive, in the ground state the spins are antiparallel between
nearest neighbor links on the same axis, as well as between
parallel links across a unit square ($\langle S^a_{i + \mu}
\rangle \sim (-1)^{i} \mathcal{S}^a_\mu$), and we will call this
state the staggered state. Since in the $J_1 - J_2$ model Eq.
\ref{model} there is no coupling between different axes along
different directions, for both cases the zero temperature ground
state of model Eq. \ref{model} has large degeneracy, because the
energy does not depend on the relative angle between
$\mathcal{S}^a_{i,x}$, $\mathcal{S}^a_{i,y}$ and
$\mathcal{S}^a_{i,z}$ $i.e.$ the ground state manifold has an
enlarged $[\mathrm{O}(N)]^3$ symmetry. However, at finite
temperature this accidental enlarged symmetry of the ground state
will be broken due to thermal fluctuation.

In this paper we will focus on the staggered spin order. Following
the magnetic field formalism of the CDM mentioned before, in order
to describe this system compactly, we introduce three flavors of
O($N$) vector field $\phi^a_\mu = S^a_{i, + \mu} \eta_i$ with $\mu
= x, y, z$, and now the constraint Eq. \ref{onconstraint} can be
rewritten concisely as \beqn \sum_\mu \nabla_\mu\phi^a_\mu = 0.
\label{phiconstraint}\eeqn Under the lattice symmetry
transformation, $\phi^a_\mu$ transforms as \beqn T_\mu &:& \ \mu
\rightarrow \mu + 1, \ \phi^a_\nu \rightarrow -\phi^a_\nu, \
(\mathrm{for} \ \mathrm{all} \ \mu, \ \nu), \cr \cr R_{\mu, s} &:&
\ \mu \rightarrow -\mu, \ \phi^a_\mu \rightarrow -\phi^a_\mu, \
\phi^a_\nu \rightarrow \phi^a_\nu, \ (\mathrm{for} \ \nu \neq
\mu), \cr\cr R_{\mu\nu} &:& \ \mu \leftrightarrow \nu, \
\phi^a_\mu \leftrightarrow \phi^a_\nu, \ (\mathrm{for} \ \nu \neq
\mu). \eeqn $T_\mu$ is the translation symmetry along $\mu$
direction, $R_{\mu,s}$ is the site centered reflection symmetry,
and $R_{\mu\nu}$ is the reflection along a diagonal direction.

\begin{figure}
\includegraphics[width=3.2in]{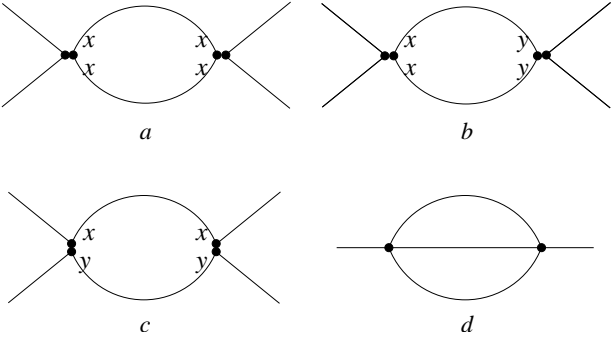}
\caption{$(a)$, $(b)$ and $(c)$, momentum-shell loop integrals
evaluated to be $A \ln (\Lambda/\tilde{\Lambda})$, $B \ln
(\Lambda/\tilde{\Lambda})$ and $C \ln (\Lambda/\tilde{\Lambda})$
respectively, with parameters given by Eq. \ref{abc}. $(d)$, the
self-energy correction that renormalizes $\gamma$ in Eq.
\ref{field}, this two loop diagram will contribute to the
renormalization of $u$, $v$ and $w$ at order $\epsilon^3$, and
hence is negligible in our calculation. } \label{fddimer}
\end{figure}

The staggered spin order corresponds to the uniform order of
$\phi^a_\mu$, and all flavors of spin vectors are ordered.
Therefore presumably $\phi^a_\mu$ are the low energy modes close
to the transition, and we can write down the following symmetry
allowed trial field theory for $\phi^a_\mu$ with softened unit
length constraint: \beqn \mathcal{F} &=& \sum_{\mu, a} \phi^a_\mu
(-\nabla^2 + r - \gamma \nabla_\mu^2)\phi^a_\mu \cr\cr &+&
\sum_{a} g (\sum_\mu \nabla_\mu\phi^a_\mu)^2 + \mathcal{F}_4
\label{field}\eeqn When we take $g \rightarrow \infty$, the
constraint Eq. \ref{phiconstraint} is effectively imposed. In Eq.
\ref{field} when $\gamma = 0$, the quadratic part of the field
theory is invariant under $\mathrm{O}(N) \times \mathrm{O}(3) $
transformation, the O(3) symmetry is a combined flavor-orbital
rotation symmetry. $\gamma$ term will break this symmetry down to
the cubic lattice symmetry and O($N$) spin symmetry. The flow of
$\gamma$ comes from the two loop self-energy correction diagram
(Fig. \ref{fddimer}$d$), and the RG flow of $\gamma$ will
contribute to the RG equation at the order of $\epsilon^3$, which
is negligible at the accuracy of our calculation if we take
$\epsilon = 4 - d$ small. Therefore $\gamma$ is a constant instead
of a scaling function in the RG equation, hereafter we will always
assume $\gamma$ is small.

$\mathcal{F}_4$ in Eq. \ref{field} includes all the symmetry
allowed quartic terms of $\phi^a_\mu$: \beqn \mathcal{F}_4 &=& u
\sum_\mu [\sum_a (\phi^a_\mu)^2]^2 + v \sum_{\mu < \nu} [\sum_a
(\phi^a_\mu)^2][\sum_b (\phi^b_\nu)^2] \cr\cr &+& w \sum_{\mu <
\nu} [\sum_a \phi^a_\mu\phi^a_\nu][\sum_b \phi^b_\mu\phi^b_\nu].
\label{f4}\eeqn The $u$ and $v$ terms are invariant under an
enlarged symmetry $[\mathrm{O}(N)]^3$, while the $w$ term breaks
this symmetry down to one single O($N$) symmetry plus lattice
symmetry. As already mentioned, the ground state manifold of model
Eq. \ref{model} has the same enlarged $[\mathrm{O}(N)]^3$
symmetry. However, the $w$ term can be induced with thermal
fluctuation through order-by-disorder mechanism \cite{henley1989},
or we can simply turn on such extra bi-quadratic term
energetically in the $J_1-J_2$ model Eq. \ref{model}. By adjusting
the ratio between $u$, $v$ and $w$ in Eq. \ref{f4}, points with
various enlarged symmetry can be found. For instance, if $w + v =
2u$, the $\mathcal{F}_4$ has a $\mathrm{O}(N)\times \mathrm{O}(3)$
symmetry, where the O(3) is the flavor-orbital combined rotation.
Just like the $J_1-J_2$ model on the square lattice
\cite{henley1989}, the quadratic coupling $\sum_a
\phi^a_\mu\phi^a_\nu$ with $\mu \neq \nu$ as well as more
complicated quartic terms like $[\sum_a \phi^a_x\phi^a_y][\sum_b
\phi^b_y\phi^b_z]$ break the reflection symmetry of the system,
and hence are forbidden.

If we just take the inverse of the Gaussian part of Eq.
\ref{field} with $r > 0$, we obtain the following correlation
function of $\phi^a_\mu$: \beqn D^{ab}_{\mu\nu} \sim
 \frac{ \delta_{ab}}{r+q^2} (\delta_{\mu\nu} - \frac{g q_\mu
q_\nu}{r + (1 + g) q^2}). \label{correlationg}\eeqn In the limit
with $g \rightarrow \infty$, the correlation function reads \beqn
\lim_{g\rightarrow \infty} D^{ab}_{\mu\nu}(\vec{q}) &=& \langle
\phi^a_\mu(\vec{q}) \phi^b_\nu(-\vec{q}) \rangle \sim
\frac{\delta_{ab}}{r+ q^2}P_{\mu\nu}, \cr\cr \ P_{\mu\nu} &=&
\delta_{\mu\nu} - \frac{q_\mu q_\nu}{q^2}.
\label{correlation}\eeqn $P_{\mu\nu}$ is a projection matrix that
projects a vector to the direction perpendicular to its momentum.
After Fourier transformation, this correlation function gives us
the $1/r^3$ power-law spin correlation of the Coulomb phase. When
$r < 0$, the vector $\phi^a_\mu$ is ordered. Interaction
$\mathcal{F}_4$ will not spontaneously generate longitudinal spin
wave. For instance, we can first keep $g$ finite at the beginning,
and calculate the leading order self-energy correction using Fig.
\ref{fddimer}$d$. It is straightforward to check that in the
result after taking the limit $g \rightarrow \infty$ the
longitudinal wave still acquires infinite kinetic energy, and the
dressed correlation function is still fully transverse.

Now a systematic renormalization group (RG) equation can be
computed with four parameters $u$, $v$, $w$ and $r$, at the
critical point $r = 0$ with the correlation function Eq.
\ref{correlation}. In the calculation an $\epsilon = 4 - d$
expansion is used, and the accuracy is kept to the first order
$\epsilon$ expansion. Based on the spirit of $\epsilon$ expansion,
all the loop integrals should be evaluated at $d = 4$, and because
of the flavor-orbital coupling imposed by the constraint Eq.
\ref{onconstraint}, we should generalize our system to four
dimension, and also increase the flavor number to $\mu = x, \ y, \
z, \ \tau$. Notice that the flavor-mixing correlation in Eq.
\ref{correlation} significantly increases the number of diagrams
that we need to evaluate. Our RG calculation is based on momentum
shell integral, since the correlation function Eq.
\ref{correlation} acquires strong momentum direction dependence,
after the momentum shell integral the logarithmic correction will
depend on the flavor of the loop integrals. For instance, the loop
diagrams in Fig. \ref{fddimer} $a$, $b$ and $c$ expanded to the
first order of $\gamma$ is evaluated as \beqn A &=& \frac{
\int_{\tilde{\Lambda} < |\vec{q}| < \Lambda} d^4q
D_{\mu\mu}(\vec{q})D_{\mu\mu}(- \vec{q})}{\int_{\tilde{\Lambda} <
|\vec{q}| < \Lambda} d^4q (1/q^2)^2 } = \frac{5}{8} -
\frac{7}{40}\gamma + O(\gamma^2),\cr\cr\cr B &=& \frac{
\int_{\tilde{\Lambda} < |\vec{q}| < \Lambda} d^4q
D_{\mu\nu}(\vec{q})D_{\mu\nu}(- \vec{q})}{\int_{\tilde{\Lambda} <
|\vec{q}| < \Lambda} d^4q (1/q^2)^2 } = \frac{1}{24} -
\frac{1}{40}\gamma + O(\gamma^2), \cr\cr\cr C &=& \frac{
\int_{\tilde{\Lambda} < |\vec{q}| < \Lambda} d^4q
D_{\mu\mu}(\vec{q})D_{\nu\nu}(- \vec{q})}{\int_{\tilde{\Lambda} <
|\vec{q}| < \Lambda} d^4q (1/q^2)^2 } = \frac{13}{24} -
\frac{23}{120}\gamma + O(\gamma^2). \cr \label{abc}\eeqn For
arbitrary $N$, the full coupled RG equation at the first order
$\epsilon$ expansion reads \beqn \frac{d u }{d\ln l} &=& \epsilon
u - 8A(8 + N)u^2 - 6N(A + 2B)v^2 \cr\cr &-& 6(A + 2B)w^2 - 24(2B +
BN)uv \cr\cr &-& 12 (A + 2B)vw - 72 B uw, \cr\cr \frac{d v }{d\ln
l} &=& \epsilon v - 16B (4 + N)u^2 - 4 (2AN + 7BN + 4C)v^2 \cr\cr
&-& 4Cw^2 - 16(2A + AN + 2BN + 4B)uv \cr\cr &-& 8(2A + 9B)vw -
16(A + 6B)uw, \cr\cr \frac{d w }{d\ln l} &=& \epsilon w - 64 B u^2
- 16 Bv^2 \cr\cr &-& 4(2A + 10 B + BN + 2 C + CN)w^2 \cr\cr &-&
16(B + 2C)vw - 32 Auw, \cr\cr \frac{dr}{d\ln l} &=& 2 r - 8(2A +
AN + 6B + 3BN)ur \cr\cr &-& 12(AN + 3BN)vr - 12(A + 3B)wr.
\label{staggeredRG} \eeqn A similar set of recursion relations of
quartic interaction terms were computed in a different context in
Ref. \cite{aharony1975}.

Let us first discuss the solution of this RG equation with $\gamma
= 0$. Solving this equation at $r = 0$ with number $A, \ B, \ C$
given by Eq. \ref{abc}, we find eight fixed points, with one
stable fixed point for $N \geq N_c = 70$, while for any $N < N_c$
only instable fixed points are found. The analytical expression of
the stable fixed point as a function of $N$ with $N > N_c$ can be
straightforwardly obtained by solving Eq. \ref{staggeredRG}, but
the result is rather lengthy. Instead, we will analyze the
solution of Eq. \ref{staggeredRG} with an expansion of $1/N$. For
instance the stable fixed point is located at \beqn u_\ast &=&
\frac{17\epsilon}{84N} -\frac{1139\epsilon}{441 N^2} +
O(\frac{\epsilon}{N^3}), \cr\cr \ v_\ast &=& - \frac{\epsilon}{42
N} - \frac{1964\epsilon}{1323N^2} + O(\frac{\epsilon}{N^3}),
\cr\cr w_\ast &=& \frac{3\epsilon}{7N} - \frac{4870\epsilon}{1323
N^2} + O(\frac{\epsilon}{N^3}), \label{fixedpoint}\eeqn Since now
$v_\ast + w_\ast = 2u_\ast$, this fixed point has the enlarged
$\mathrm{O}(N)\times \mathrm{O}(d)$ symmetry mentioned before.
Close to the stable fixed point, the three eigenvectors of the RG
flow have scaling dimensions \beqn \Delta_1 &=& - \epsilon +
O(\frac{\epsilon}{N^2}), \cr\cr \Delta_2 &=& - \epsilon +
\frac{4448\epsilon}{567N} + O(\frac{\epsilon}{N^2}), \cr\cr
\Delta_3 &=& - \epsilon + \frac{24950\epsilon}{567N} +
O(\frac{\epsilon}{N^2}). \label{dimensions}\eeqn

At the stable fixed point Eq. \ref{fixedpoint}, $r$ is the only
relevant perturbation, and plugging the fixed point values Eq.
\ref{fixedpoint} back to the RG equation, we obtain the critical
scaling dimension \beqn [r] = \frac{1}{\nu} = 2 - \epsilon +
\frac{158\epsilon}{7N} + O(\frac{\epsilon}{N^2}) \eeqn Since at
the ground state all three flavors of spin vectors are ordered, in
the field theory $\mathcal{F}_4$, $v$ should be smaller than $2u$,
which is well consistent with the stable fixed point in Eq.
\ref{fixedpoint} with negative $v_\ast$. This fixed point has
positive $w_\ast$, which favors noncollinear alignment between
spins on different axes. Therefore the transition between Coulomb
and noncollinear staggered state has a better chance to be
described by this fixed point.

Notice that had we included the anisotropic velocity $\gamma$ into
account, its leading RG flow will be at order of $\epsilon^2$, and
the flow of $\gamma$ will contribute to the RG flow of $u$, $v$
and $w$ at order of $\epsilon^3$, therefore it is justified to
take $\gamma$ a constant in our calculation as long as we keep
$\epsilon$ small enough. When $\gamma$ is nonzero but small, we
can solve the RG equation with $A$, $B$ and $C$ given by Eq.
\ref{abc}, and the RG flows will only change quantitatively,
although the $\mathrm{O}(N)\times \mathrm{O}(3)$ symmetry of the
stable fixed point is broken by $\gamma$. Expanded to the first
order of $\gamma$, the scaling dimensions of the three
eigenvectors of the RG equation at the stable fixed point become
$\Delta_1 = - \epsilon$, $\Delta_2 = - \epsilon +
\frac{4448\epsilon}{567N} +
\frac{2849936\epsilon\gamma}{3988845N}$, $\Delta_3 = - \epsilon +
\frac{24950\epsilon}{567N} - \frac{30088528 \epsilon
\gamma}{27921915N}$, and the scaling dimension of $r$ becomes $[r]
= \frac{1}{\nu} = 2 - \epsilon + \frac{158\epsilon}{7N} -
\frac{1032 \epsilon\gamma }{1715 N}$.

\begin{figure}
\includegraphics[width=3.4in]{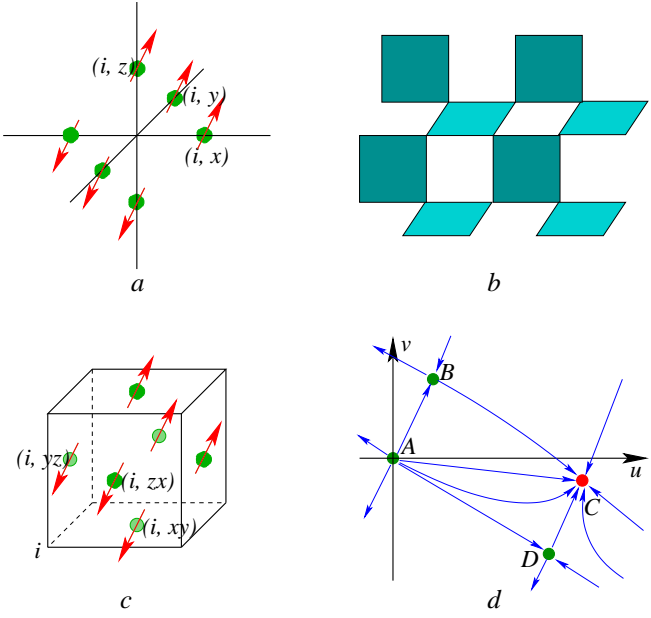}
\caption{$(a)$, the lattice structure of the O($N$) spin-ice
model. $(b)$, one of the plaquette configurations of CPM-2, with
every site shared by two plaquettes. $(c)$, lattice structure for
the O($N$) spin-plaquette model, each unit square face of the
cubic lattice is occupied by an O($N$) vector, and the constraint
is that the sum of all the O($N$) vectors on the twelve square
faces shared by site $i$ is zero. $(d)$, the RG flow of Eq.
\ref{staggeredRG} in the large-$N$ limit, $C$ is the stable fixed
point, with enlarged $\mathrm{O}(N)\times \mathrm{O}(d)$
symmetry.} \label{dimeron}
\end{figure}

If we took the limit $N \rightarrow \infty$ in the physical
system, we only need to keep the terms linear with $N$ in Eq.
\ref{staggeredRG}, and now the equation becomes precise even with
$\epsilon = 1$. In this case the RG flow of $w$ is decoupled from
$u$ and $v$, hence four of the eight fixed points have $w_\ast =
0$, and all the others have $w_\ast = 3/(7N)$. The RG flow diagram
for $u$ and $v$ with $w = w_\ast = 3/(7N)$ in the large-$N$ limit
is depicted in Fig. \ref{dimeron}$d$.

As we promised in the beginning of this paper, we should discuss
the applicability of the constrained GL formalism discussed in
this paper to the CDM-3, which corresponds to the case with $N =
1$. In our GL formalism, in the ordered phase, the power law
spin-spin correlation still persists if the long range correlation
is subtracted. For instance, the fluctuation $\delta\phi_\mu =
\phi_\mu - \langle \phi_\mu \rangle$ is still subject to the
constraint $ \sum_\mu \nabla_\mu \delta \phi_\mu = 0$, therefore
although the fluctuation is gapped, it still leads to the $1/r^3$
power-law correlation. But in CDM, the ordered phase only has
short range connected dimer correlation on top of the long range
order, which can be checked with a low temperature expansion of
CDM \cite{balentsnote}. Like what was discussed in the
introduction, the key property of the case for $N = 1$ is that,
the spins only take discrete value $\pm 1$, the vertex operator
like Eq. \ref{vertexmatter} in the dual field theory in terms of
vector potential can drive the system to a phase with short range
connected correlation through a Higgs transition. The effect of
the vertex operator was missing in our GL formalism.

We can solve Eq. \ref{staggeredRG} with $N = 1$, where $v$ and $w$
terms are identical. In this case in addition to the trivial
Gaussian fixed point, there is only one other fixed point at
$v_\ast = 2u_\ast = \epsilon/34$ with O(3) flavor-space combined
rotation symmetry, which is the same fixed point as the
ferromagnetic transition with dipolar interaction
\cite{dipolar1,dipolar2}. In 3d space, the dipolar interaction
also projects a spin wave to its transverse direction. The dipolar
fixed point is instable against the O(3) to cubic symmetry
breaking, therefore when $N = 1$ our first order $\epsilon$
expansion predicts a first order transition. Based on the
discussion in the previous paragraph, one possible scenario for
the CDM-3 with staggered ground state is that, if we lower the
temperature from the Coulomb phase, after the first order
transition of $\phi^a$, there has to be another ``Higgs" like
phase transition that destroys the power-law connected
correlation. Or there can be one single strong first order
transition that connects the Coulomb phase and staggered dimer
crystal directly.

\subsection{C, O($N$) Spin-Ice with Easy Plane Anisotropy}

Model Eq. \ref{model} is invariant under cubic lattice symmetry,
and we can certainly turn on various anisotropies to this model,
like what was studied in the CDM-1 model \cite{balentsdimer}. For
instance, let us modify the model Eq. \ref{model} slightly: \beqn
E &=& \sum_{i,\mu, a} J_{1, \mu} S^a_{i -\mu, \mu}S^a_{i,\mu} +
\sum_{i,a}\sum_{\mu\neq\nu} J_{2, \mu\nu} S^a_{i,\mu}S^a_{i +
\nu,\mu}. \label{model2}\eeqn If $J_{1,x} = J_{1,y} > J_{1,z} >
0$, and $J_{2,xy} = J_{2,xz} = J_{2,yx} = J_{2,yz} > J_{2,zx} =
J_{2,zy} > 0$, the O($N$) spin vectors in the $xy$ plane,
$S^a_{i,x}$ and $S^a_{i, y}$ have a stronger tendency to order
compared with $S^a_{i,z}$. Therefore when we lower the temperature
from the high temperature algebraic phase, the O($N$) vectors in
$xy$ plane are expected to order first at critical temperature
$T_{c1}$. In the field theory close to $T_{c1}$, the anisotropy
can be described by an extra mass term for $\sum_a m^2
(\phi^a_z)^2$ in the free energy, which is clearly a relevant
perturbation at the critical point $r = 0$: \beqn \mathcal{F} &=&
\sum_{\mu, a} \phi^a_\mu (-\nabla^2 + r)\phi^a_\mu + \sum_a m^2
(\phi^a_z)^2 \cr\cr &+& \sum_{a} g (\sum_\mu
\nabla_\mu\phi^a_\mu)^2 + \mathcal{F}_4. \label{field2} \eeqn

In the equation above we have taken $\gamma = 0$ for simplicity.
To calculate the RG equations for $\mathcal{F}_4$, we still need
to increase the dimension and flavor number to four with $\mu = x,
\ y, \ z, \ \tau$, and the anisotropy of the generalized system
will prefer the O($N$) vectors to order on three of the four axes.
At the critical point $r = 0$, due to the relevance of the extra
mass term, we can safely take $m \rightarrow \infty$, and the
correlation function between $\phi^a_\mu$ reads \beqn \lim_{g, m
\rightarrow \infty, r \rightarrow 0} && D^{ab}_{\mu\nu}(\vec{q}) =
\langle \phi^a_\mu(\vec{q})\phi^b_{\nu}(-\vec{q}) \rangle \sim
\frac{\delta_{ab}}{q^2}Q_{\mu\nu}, \cr\cr Q_{\mu\nu} &=&
\delta_{\mu\nu} - \frac{q_\mu q_\nu}{\mathrm{\mathbf{q}}^2}, \
\mu, \nu \neq \tau, \ \mathrm{\mathbf{q}}^2 = q_x^2 + q_y^2 +
q_\tau^2 , \cr\cr Q_{\mu\nu} &=& 0, \ \mu \ \mathrm{or} \ \nu =
\tau. \label{correlation2}\eeqn Using the correlation function Eq.
\ref{correlation2}, the RG equation at the critical point reads
\beqn \frac{d u }{d\ln l} &=& \epsilon u - 8A(8 + N)u^2 - 4N(A +
B)v^2 \cr\cr &-& 4(A + B)w^2 - 4B(8 + 4N)uv \cr\cr &-& 8 (A + B)vw
- 48 B uw, \cr\cr \frac{d v }{d\ln l} &=& \epsilon v - 16B (4 +
N)u^2 - 4 (AN + 3BN + 4C)v^2 \cr\cr &-& 4Cw^2 - 16(2A + AN + BN +
2B)uv \cr\cr &-& 8(A + 5B)vw - 16(A + 3B)uw, \cr\cr \frac{d w
}{d\ln l} &=& \epsilon w - 64 B u^2 - 16 Bv^2 \cr\cr &-& 4(A + 6 B
+ BN + 2 C + CN)w^2 \cr\cr &-& 16(B + 2C)vw - 32 Auw, \cr\cr
\frac{dr}{d\ln l} &=& 2 r - 8(2A + AN + 4B + 2BN)ur \cr\cr &-&
8(AN + 2BN)vr - 8(A + 2B)wr. \label{staggeredRG2} \eeqn Now $A =
8/15$, $B = 1/15$, $C = 2/5$, which is different from the
isotropic case, due to the different form of the correlation
functions. Solving this equation at $r = 0$, we again find a
stable fixed point for large enough $N$. Expanded to the order of
$\epsilon/N^2$, the stable fixed point is located at \beqn u_\ast
&=& \frac{27\epsilon}{112N} -\frac{1431\epsilon}{490 N^2} +
O(\frac{\epsilon}{N^3}), \cr\cr \ v_\ast &=& - \frac{3\epsilon}{56
N} - \frac{879\epsilon}{490N^2} + O(\frac{\epsilon}{N^3}), \cr\cr
w_\ast &=& \frac{15\epsilon}{28N} - \frac{1983\epsilon}{490 N^2} +
O(\frac{\epsilon}{N^3}), \label{fixedpoint2}\eeqn Again, since
$v_\ast + w_\ast = 2u_\ast$, this fixed point has an enlarged
$\mathrm{O}(N)\times \mathrm{O}(d-1)$ symmetry. Close to the
stable fixed point, the three eigenvectors of the RG flow have
scaling dimensions \beqn \Delta_1 &=& - \epsilon +
O(\frac{\epsilon}{N^2}), \cr\cr \Delta_2 &=& - \epsilon +
\frac{942\epsilon}{25N} + O(\frac{\epsilon}{N^2}), \cr\cr \Delta_3
&=& - \epsilon + \frac{1104\epsilon}{175N} +
O(\frac{\epsilon}{N^2}). \label{dimensions2}\eeqn $\Delta_2$ is
the largest scaling dimension, and according to Eq.
\ref{staggeredRG2} the critical $N$ is $N_c \sim 59$, below which
the stable fixed point disappears. Plug the fixed point value Eq.
\ref{fixedpoint2} back to the last RG equation in Eq.
\ref{staggeredRG2}, we obtain the scaling dimension of $r$ at the
fixed point: \beqn [r] = \frac{1}{\nu} = 2 - \epsilon +
\frac{138\epsilon}{7N} + O(\frac{\epsilon}{N^2}). \eeqn

If we keep lowering the temperature after the order of $\phi^a_x$
and $\phi^a_y$, then eventually $\phi^a_z$ will also order at
temperature $T_{c2} < T_{c1}$. After the order of $\phi^a_x$ and
$\phi^a_y$, the O($N$) symmetry is broken down to its subgroup.
Let us first assume $w < 0$ in $\mathcal{F}_4$ $i.e.$ the three
flavors of O($N$) vectors are collinear with each other in the low
temperature ordered phase. Let us assume that in the intermediate
phase the expectation value $ \langle \phi^a_x \rangle \sim
\langle \phi^a_y \rangle \sim (1, 0, \cdots 0)$, now the O($N$)
symmetry is broken down to O($N - 1$) symmetry generated by Lie
algebra $\Gamma_{ab}$ with $a,b = 2, \cdots N$. This symmetry
breaking implies that the degeneracy of the $N$ components of
$\phi^a_z$ will be lifted at $T_{c2}$. Because $w < 0$, the
nonzero expectation value of $\phi^1_x$ and $\phi^1_y$ will prefer
$\phi^1_z$ to order next at lower temperature, which is
essentially an Ising transition with order parameter $\phi^1_z$.
From now on we will denote $\phi^1_z$ as $\phi_z$. $\phi_z$ will
couple to the gapped fluctuations $\phi_x = \phi^1_x - \langle
\phi^1_x \rangle $ and $\phi_y = \phi^1_y - \langle \phi^1_y
\rangle $ through the constraint, and the entire low energy field
theory at $T_{c2}$ reads \beqn \mathcal{F} &=&
\phi_z(-\nabla^2)\phi_z + \sum_{\mu = x, y}\phi_\mu(- \nabla^2 +
m^2)\phi_\mu \cr\cr &+& g (\sum_\mu \nabla_\mu \phi_\mu)^2 +
O(\phi_z^4). \eeqn After taking the limit $g \rightarrow \infty$,
the critical correlation function of $\phi_z$ reads \beqn \lim_{g
\rightarrow \infty} D_{zz}(\vec{q}) \sim \frac{1}{m^2
\frac{q_z^2}{q_x^2 + q_y^2} + q_x^2+ q_y^2 + \cdots}. \label{z2}
\eeqn Therefore this transition at $T_{c2}$ is effectively a $z =
2$ transition, with scaling dimension $[q_z] = 2[q_x] = 2[q_y] =
2$. Now the total effective dimension is 4, and the $(\phi_z)^4$
term is a marginally irrelevant perturbation, therefore this
transition is a mean field transition with logarithmic
corrections. Notice that the Goldstone modes after the O($N$) to
O($N - 1$) symmetry breaking are harmless to this transition. The
Goldstone mode can be described by $\delta \phi^a_x \sim \delta
\phi^a_y \sim (0, \pi^2, \cdots \pi^N)$ that forms a vector
representation of O($N - 1$), and in order to guarantee the
gaplessness of the Goldstone modes, close to $T_{c2}$ the
following coupling between $\phi_z$ and $\vec{\pi}$ is the lowest
order coupling that is allowed: \beqn \mathcal{F}_{goldstone} \sim
\sum_\mu \phi_z^2 (\nabla_\mu \vec{\pi})^2. \eeqn This term only
generates irrelevant perturbations for $\phi_z$ at the transition.
Couplings like $\phi_z^2 (\vec{\pi})^2$ is forbidden due to its
ability to renormalize the mass of Goldstone mode $\vec{\pi}$.

If $w > 0$ in $\mathcal{F}_4$, then the three flavors of O($N$)
vectors are perpendicular to each other in the low temperature
ordered phase. 
Let us assume that in the intermediate phase between $T_{c1}$ and
$T_{c2}$ the expectation values $\langle \phi^a_x \rangle \sim (1,
0, \cdots 0)$ and $\langle \phi^a_y \rangle \sim (0, 1, 0 \cdots
0)$. Now the O($N$) symmetry is broken down to the O($N - 2$)
symmetry generated by $\Gamma_{ab}$ with $a,b = 3,\cdots N$. Due
to the presence of $w$ term in $\mathcal{F}_4$, at $T_{c2}$ the
order parameters should be $\phi^a_z$ with $a = 3 \cdots N$, which
forms a vector representation of O($N - 2$). In the intermediate
phase there are in total $2N - 3$ Goldstone modes, they are
$\pi^a_x = (0, 0, \pi^3_x, \cdots \pi^N_x)$ that corresponds to
the O($N$) Lie algebra elements $\Gamma_{1a}$ with $a = 3, \cdots
N$; $\pi^a_y = (0, 0, \pi^3_y \cdots \pi^N_y)$ that correspond to
the the O($N$) Lie algebra elements $\Gamma_{2a}$ with $a = 3,
\cdots N$; plus a rotation mode between $\phi^1_x$ and $\phi^2_x$,
denoted as $\Gamma_{12}$. Again the order parameter $\phi^a_z$
will couple to these Goldstone modes through the constraint. The
field theory close to $T_{c2}$ reads \beqn \mathcal{F} &=&
\sum_{\mu = x, y}\sum_{a = 3}^N \pi^a_\mu(-\nabla^2)\pi^a_\mu +
\sum_{a = 3}^N \phi^a_z(- \nabla^2 + r)\phi^a_z \cr\cr &+& \sum_{a
= 3}^N g (\nabla_z \phi^a_z + \nabla_x\pi^a_x + \nabla_y\pi^a_y)^2
+ O[(\phi_z^a)^4]. \eeqn Notice that the interaction between
Goldstone modes $\pi^a_\mu$ have to be irrelevant to guarantee
their gaplessness. The universality class of this transition at $r
= 0$ can be calculated using the correlation function of
$\phi^a_z$ after taking the limit $g \rightarrow \infty$: \beqn
\lim_{g \rightarrow \infty, r \rightarrow 0} D^{ab}_{zz}(\vec{q})
\sim \frac{(q_x^2 + q_y^2)}{(q_x^2 + q_y^2 + q_z^2)^2}\delta_{ab}.
\eeqn The first order $\epsilon$ expansion leads to the following
scaling dimension and critical exponent: \beqn [r] = \frac{1}{\nu}
= 2 - \frac{N-1}{N+5}\epsilon, \eeqn which is identical to the 3d
O($N - 2$) Wilson-Fisher (WF) fixed point, although higher order
expansions may deviate from the WF fixed point. Therefore our
formalism implies that with large enough $N$, the easy plane
anisotropy will split the transition discussed in the previous
subsection to two second order transitions.

\subsection{D, O($N$) Spin-Ice with Easy Axis Anisotropy}

In model Eq. \ref{model2}, if we make the following choice of
parameters: $J_{1,z} > J_{1,x} = J_{1,y} > 0$, $J_{2,zx} =
J_{2,zy} > J_{2,xy} = J_{2, xz} = J_{2,yx} = J_{2,yz}$, then when
we lower the temperature from the algebraic phase, the O($N$)
vectors along the $z$ axis will order first at temperature
$T_{c1}$. Similar to the previous section, this easy axis
anisotropy can be described by an extra mass gap $m$ for both
$\phi^a_x$ and $\phi^b_y$ modes in the field theory Eq.
\ref{field}, and $\phi^a_z$ becomes the only order parameter at
low energy. However, now we can no longer take the limit $m
\rightarrow \infty$, because with this limit all the correlation
functions will vanish. If we keep $m$ finite, the correlation
function of $\phi^a_z$ takes a similar form as Eq. \ref{z2}: \beqn
\lim_{g \rightarrow \infty} D^{ab}_{zz}(\vec{q}) \sim
\frac{\delta_{ab}}{m^2 \frac{q_z^2}{q_x^2 + q_y^2} + q_x^2+ q_y^2
+ \cdots}, \label{z22}\eeqn therefore this transition is again an
effective a $z = 2$ mean field transition. A similar situation
with easy axis anisotropy has been studied in Ref.
\cite{anisotropy}, although there the easy axis was along the
diagonal direction.

If the temperature is lowered even more from $T_{c1}$, then at
$T_{c2}$ the spin vectors within $xy$ planes will order next.
Again the nature of this transition would depend on the sign of
$w$ in $\mathcal{F}_4$. If $w < 0$ $i.e.$ the three flavors of
O($N$) vectors order collinearly, then if $\phi^a_z$ orders along
direction $\langle\phi^a_z\rangle \sim (1, 0 \cdots 0)$, at
temperature $T_{c2}$ $\phi^1_x$ and $\phi^1_y$ will become the
critical modes. $\phi^1_x$ and $\phi^1_y$ are coupled to the
gapped fluctuation $\phi_z = \phi^1_z - \langle \phi^1_z \rangle$,
and the low energy field theory describing $\phi_x$ and $\phi_y$
is identical to Eq. \ref{field2} with $N = 1$. By solving the RG
equation Eq. \ref{staggeredRG2} with $N = 1$, no stable fixed
point is found.

If $w > 0$, the critical modes at $T_{c2}$ is $\phi^a_{x}$ and
$\phi^a_y$ with $a = 2, \cdots N$, which form vector
representations of O($N - 1$). These two critical modes are
coupled to gapless Goldstone mode $\pi^a_z = (0, \pi^2_z, \cdots
\pi^N_z)$, with field theory \beqn \mathcal{F} &=& \sum_{\mu = x,
y}\sum_{a = 2}^N \phi^a_\mu(-\nabla^2 + r)\phi^a_\mu + \sum_{a =
2}^N K \pi^a_z(- \nabla^2 )\pi^a_z \cr\cr &+& \sum_{a = 3}^N g
(\nabla_z \pi^a_z + \nabla_x\phi^a_x + \nabla_y\phi^a_y)^2 +
\mathcal{F}_4(\phi^a_x, \phi_y^a). \eeqn Now $\mathcal{F}_4$ only
involves two flavors of O($N - 1$) vectors $\phi^a_x$ and
$\phi^a_y$. Number $K$ is in general not 1 because the now there
is no symmetry that can transform $\phi^a_x$, $\phi^a_y$ and
$\pi^a_z$. Again, the renormalization of $K$ will be an order
$\epsilon^2$ effect, therefore we can take $K$ as a constant. The
RG equation for $\mathcal{F}_4$ can be calculated in the same
manner as we did before. If we take $K = 1$, then the RG equation
for $\mathcal{F}_4$ takes the same form as Eq. \ref{staggeredRG2}
but with $A = 5/8$, $B = 1/24$, $C = 13/24$. With large enough $N$
there is still a stable fixed point, and the answer is
qualitatively unchanged when $K$ deviates from 1 slightly.

\section{III, O($N$) Spin-Plaquette Model}

\subsection{A, Phase Transition with General $N$}

Now we switch the gear to the less well studied spin-plaquette
model, where the physical quantities are defined on the unit
square faces of the cubic lattice instead of the links. In Ref.
\cite{xuwu2008,pankov2007}, a model of this type was studied, with
every site of the cubic lattice connects to precisely one filled
plaquette. Physically the filled plaquette is a SU(4) singlet
formed with four SU(4) fundamental fermions on four corners of the
plaquette. This model is denoted as CPM-1. In this section we will
study an O($N$) generalization of this model. We define an O($N$)
spin vector $S^a_{i, \mu\nu}$ on each unit face of the cubic
lattice, and impose the constraint \beqn \sum_{\mu,\nu = x, y, z}
S^a_{i,\mu\nu} + S^a_{i - \mu, \mu\nu} + S^a_{i - \nu, \mu\nu} +
S^a_{i - \mu - \nu, \mu\nu} = 0. \eeqn In addition to this
constraint, we consider the following Hamiltonian: \beqn E =
\sum_{i, \mu, \nu} J_1 S^a_{i, \mu\nu}S^a_{i + \mu, \mu \nu} + J_2
S^a_{i, \mu\nu} S^a_{i + \rho, \mu\nu}, \rho \neq \mu, \nu.
\label{model3}\eeqn When $J_1 > 0$, $J_2 > 0$, the ground state of
this Hamiltonian has staggered order with nonzero $\langle
\phi^a_{i, \mu\nu} \rangle \sim (-1)^i S^a_{i, \mu\nu}$ with $\mu,
\nu = +x, + y, +z$. $\phi^a_{\mu\nu}$ is a symmetric tensor with
$\mu \neq \nu$, which has only three independent flavors. Under
discrete cubic lattice symmetry, tensor field $\phi^a_{\mu\nu}$
transforms as \beqn T_\mu &:& \ i \rightarrow i + \mu, \
\phi^a_{\nu\rho} \rightarrow -\phi^a_{\nu\rho}, \ (\mathrm{for} \
\mathrm{all} \ \mu, \ \nu, \ \rho), \cr\cr R_{\mu, s} &:& \ \mu
\rightarrow -\mu, \ \phi^a_{\mu\nu} \rightarrow -\phi^a_{\mu\nu},
\ \phi^a_{\nu\rho} \rightarrow \phi^a_{\nu\rho}, \ (\mathrm{for} \
\nu, \rho \neq \mu), \cr\cr R_{\mu\nu} &:& \ \mu \leftrightarrow
\nu, \ \phi^a_{\mu\rho} \leftrightarrow \phi^a_{\nu\rho}, \
(\mathrm{for} \ \nu \neq \mu). \eeqn

Just like the spin-ice model Eq. \ref{model}, the ground state of
model Eq. \ref{model3} also has enlarged symmetry
$[\mathrm{O}(N)]^3$ $i.e.$ the spins on different planes will
order independently. Since $\phi^a_{\mu\nu}$ is subject to the
constraint Eq. \ref{phi2constraint}, we can start with the
following low energy field theory close to the transition: \beqn
\mathcal{F} &=& \sum_{\mu \neq \nu, a} \phi^a_{\mu\nu} (-\nabla^2
+ r)\phi^a_{\mu\nu} + \sum_{a} g (\sum_{\mu,\nu}
\nabla_\mu\nabla_\nu \phi^a_{\mu\nu})^2 \cr\cr &+& \mathcal{F}_4,
\label{field3}\eeqn and $\mathcal{F}_4$ is all the quartic terms
allowed by symmetry: \beqn \mathcal{F}_4 &=& u \sum_{\mu < \nu}
[\sum_a (\phi^a_{\mu\nu})^2]^2 + v \sum_{\nu < \rho} [\sum_a
(\phi^a_{\mu\nu})^2][\sum_b (\phi^b_{\mu\rho})^2] \cr\cr &+& w
\sum_{\nu \neq \rho} [\sum_a
\phi^a_{\mu\nu}\phi^a_{\mu\rho}][\sum_b
\phi^b_{\mu\nu}\phi^b_{\mu\rho}]. \label{f4p}\eeqn Under the limit
$g \rightarrow \infty$, the constraint Eq. \ref{phi2constraint} is
effectively imposed, and the correlation function with $r > 0$ is:
\beqn \lim_{g \rightarrow \infty} D^{ab}_{\mu\nu, \rho\sigma} &=&
\langle \phi^a_{\mu\nu}(\vec{q})\phi^b_{\rho\sigma}(- \vec{q})
\rangle \sim \frac{\delta_{ab}}{r + q^2}R_{\mu\nu\rho\sigma},
\cr\cr R_{\mu\nu\rho\sigma} &=&
(\delta_{\mu\rho}\delta_{\nu\sigma} +
\delta_{\mu\sigma}\delta_{\nu\rho} \cr\cr &-& \frac{q_\mu q_\nu
q_\rho q_\sigma}{q_x^2q_y^2 + q_y^2 q_z^2 + q_x^2q_z^2}).
\label{correlation3}\eeqn Notice that in Eq. \ref{field3}, some
extra flavor-orbital coupling quadratic terms are allowed by
symmetry, such as \beqn \mathcal{F}_2 \sim \sum_a \nabla_x
\phi^a_{yz}\nabla_y\phi^a_{xz} + \nabla_y
\phi^a_{zx}\nabla_z\phi^a_{xy} + \nabla_z
\phi^a_{xy}\nabla_x\phi^a_{yz}. \eeqn Just like the $\gamma$ term
in Eq. \ref{field}, this term will not gain any renormalization at
the one-loop calculation $i.e.$ its RG flow will only affect the
RG equation of $u$, $v$ and $w$ at the order of $\epsilon^3$. A
small perturbation of this term will not change the RG flow
qualitatively, we will take this term to be zero hereafter for
simplicity.

To calculate the RG equation at the critical point $r = 0$, in
principle we need to generalize the model to 4d. However, there is
no ideal way to make this generalization. For instance, if we
define O($N$) vectors on the 2d faces of a 4d lattice, and $\mu$
takes $x, \ y, \ z, \ \tau$, at 4d there are two extra quartic
terms in addition to $\mathcal{F}_4$ in Eq. \ref{f4p}: \beqn
\mathcal{F}_{4a} &\sim& [\sum_a\phi^a_{xy}\phi^a_{yz}][\sum_b
\phi^b_{z\tau}\phi^b_{\tau x}] + \cdots
\cr\cr \mathcal{F}_{4b} &\sim&
[\sum_a\phi^a_{xy}\phi^a_{z\tau}][\sum_b \phi^b_{yz}\phi^b_{\tau
x}] + \cdots \eeqn These two extra terms will make the RG equation
much more complicated than the actual 3d case. For instance, since
these two terms are allowed by symmetry, they will be generated
under RG flows even if we take them to be zero at the beginning.
Therefore in this section we will just evaluate the loop integrals
at 3d.

The RG equation takes exactly the same form as Eq.
\ref{staggeredRG2}, while now $A = 5/9$, $B = 1/18$, $C = 7/18$.
With large enough $N$, there is a stable fixed point located at
\beqn u_\ast &=& \frac{11\epsilon}{48N}
-\frac{29579\epsilon}{10368 N^2} + O(\frac{\epsilon}{N^3}), \cr\cr
\ v_\ast &=& - \frac{\epsilon}{24 N} -
\frac{19039\epsilon}{10368N^2} + O(\frac{\epsilon}{N^3}), \cr\cr
w_\ast &=& \frac{9\epsilon}{16N} - \frac{22171\epsilon}{5184 N^2}
+ O(\frac{\epsilon}{N^3}), \label{fixedpoint3}\eeqn Close to the
stable fixed point, the three eigenvectors of the RG flow have
scaling dimensions \beqn \Delta_1 &=& - \epsilon +
O(\frac{\epsilon}{N^2}), \cr\cr \Delta_2 &=& - \epsilon +
\frac{6.80 \epsilon}{N} + O(\frac{\epsilon}{N^2}), \cr\cr \Delta_3
&=& - \epsilon + \frac{37.5\epsilon}{N} + O(\frac{\epsilon}{N^2}).
\label{dimensions3}\eeqn Plug the fixed point value Eq.
\ref{fixedpoint2} back to the last RG equation in Eq.
\ref{staggeredRG2}, we obtain the scaling dimension of $r$ at the
fixed point: \beqn [r] = \frac{1}{\nu} = 2 - \epsilon +
\frac{2113\epsilon}{108N} + O(\frac{\epsilon}{N^2}) \eeqn If $N =
1$, no stable fixed point is found. The case $N = 1$ is equivalent
to CPM-6, which is much less studied compared with the CDM-$m$. We
will discuss this model in the next subsection.

\subsection{B, Duality and $N = 1$}

Besides taking the limit $g \rightarrow \infty$ in the Gaussian
field theory Eq. \ref{field3}, the correlation function in Eq.
\ref{correlation3} can be obtained by other means. Let us solve
the constraint Eq. \ref{phi2constraint} with softened unit length
constraint of $\phi^a_{\mu\nu}$ by defining the three flavors of
O($N$) height field $h^a_\mu$ on the cubic center $\bar{i}$ of the
lattice: \beqn \phi^a_{xy} &=& \nabla_z (h^a_x - h^a_y), \cr\cr
\phi^a_{zy} &=& \nabla_x (h^a_z - h^a_y), \cr\cr \phi^a_{zx} &=&
\nabla_y (h^a_z - h^a_x) . \label{height} \eeqn Then height field
$h^a_\mu$ plays the same role as vector potential $A_\mu$ in the
usual dimer model. Now in terms of the height field, the algebraic
phase can be described by the following Gaussian field theory
\beqn \mathcal{F} &=&  \sum_a \frac{K}{2} [\nabla_z(h^a_x -
h^a_y)]^2 + [\nabla_y(h^a_z - h^a_x)]^2 \cr\cr &+& [\nabla_x(h^a_y
- h^a_z)]^2 + \cdots \label{heightfield}\eeqn This Gaussian field
theory takes a similar form as the Khaliullin model describing the
orbital degrees of freedom \cite{khaliullin2000}, after taking the
spin-wave expansion. Diagonalizing this Gaussian field theory, we
obtain two eigenmodes describing the fluctuations of the height
vector field: \beqn \omega_1^2 \sim \frac{q_x^2q_y^2 + q_y^2q_z^2
+ q_z^2q_x^2}{q^2}, \ \ \ \omega_2^2 \sim q^2. \label{modes}\eeqn
Interestingly $\omega_1$ vanishes along each coordinate axis in
the momentum space. This height field theory Eq. \ref{heightfield}
is invariant under the following symmetry transformation: \beqn
h^a_x \rightarrow h^a_x + f_1(x) + \varphi(x,y,z), \cr\cr h^a_y
\rightarrow h^a_y + f_2(y) + \varphi(x,y,z), \cr\cr h^a_z
\rightarrow h^a_z + f_3(z) + \varphi(x,y,z). \label{quasilocal}
\eeqn $f_\alpha$ are functions of only one of the coordinates,
this type of quasi-local symmetry comes from the definition of the
height field Eq. \ref{height}, and hence does not depend on the
detailed form of the Hamiltonian. This quasi-local symmetry is
responsible for the line of nodes of the eigenmodes in Eq.
\ref{modes}. Using the Gaussian free energy, and the height
representation of $\phi^a_{\mu\nu}$ in Eq. \ref{height}, we can
reproduce the correlation function Eq. \ref{correlation3}.

These dualities are particularly useful for $N = 1$, which is
equivalent to CPM-6. Since in this case $\phi_{\mu\nu}$ takes
discrete values, then $h_\mu$ are also discrete. Therefore in the
dual field theory Eq. \ref{heightfield} for $N = 1$ we should also
consider the vertex operators like we introduced for the CDM, in
section II-A. For convenience, let us make the following standard
modification of our description of the system: we will allow
$\phi_{\mu\nu}$ to take all the half-integer values, and turn on
classical Hamiltonian on the lattice $E = \sum_{i,\mu \neq \nu}
U(\phi_{i, \mu\nu})^2$, on top of the constraint $\sum_{\mu\neq
\nu} \nabla_\mu\nabla_\nu\phi_{\mu\nu} = 0$. When $U$ is large,
effectively on every unit face $\phi_{i,\mu\nu}$ can only take two
values, which is the same as the CPM-6. In this way, even though
we increased the total configurations of $\phi_{\mu\nu}$, the low
energy configurations of $\phi_{\mu\nu}$ is still identical to the
CPM-6. Now the dual field theory of the algebraic phase reads
\beqn \mathcal{F} &=& \frac{K}{2} [\nabla_z(h_x - h_y)]^2 +
[\nabla_y(h_z - h_x)]^2 \cr\cr &+& [\nabla_x(h_y - h_z)]^2.
\label{heightfield2} \eeqn In order to make sure $\phi_{\mu\nu}$
take half-integer values, based on the definite Eq. \ref{height},
$h_\mu - h_\nu$ will take site-dependent integer or half-integer
values on the 3d dual cubic lattice: \beqn h_x - h_y &\in&
\mathbb{Z} + \frac{1 - (-1)^{\bar{i}_z + \bar{i}_x}}{4}, \cr\cr
h_y - h_z &\in& \mathbb{Z} + \frac{1 - (-1)^{\bar{i}_x +
\bar{i}_y}}{4}, \cr\cr h_z - h_x &\in& \mathbb{Z} + \frac{1 -
(-1)^{\bar{i}_y + \bar{i}_z}}{4}. \eeqn Then in the dual theory
the lowest order vertex operator that do not have spatial
oscillation is \beqn \mathcal{F}_v = \sum_{\mu\neq\nu} - \alpha
\cos[4\pi(h_\mu - h_\nu)]. \label{heightvertex} \eeqn It was shown
in Ref. \cite{xuwu2008} that this vertex operator has directional
dependent algebraic correlation in the algebraic phase, with
scaling dimension proportional to $K$ in Eq. \ref{heightfield2}.
For instance, let us denote $\cos[4\pi(h_x - h_y)]$ as
$\mathcal{F}_{v,xy}$, then due to the quasi-local symmetry in Eq.
\ref{quasilocal}, two $\mathcal{F}_{v,xy}$ operators can only have
nonzero correlation when they are on the same $z$ axis: \beqn &&
\langle \mathcal{F}_{v,xy}(0,0,0)\mathcal{F}_{v,xy}(0,0,z) \rangle
\cr\cr &\sim& \exp\{ - \frac{4(2\pi)^2}{K} \int
\frac{d^3k}{(2\pi)^3} \frac{(k_x^2 + k_y^2)e^{ik_z z}}{k_x^2k_y^2
+ k_y^2k_z^2 + k_x^2k_z^2} \} \cr\cr &\sim& \frac{1}{z^{16\pi/
K}}. \label{heightcorrelation}\eeqn If the scaling dimension
$8\pi/K$ is greater than 1 $i.e.$ $K > K_{c1} = 8\pi$, the vertex
operator Eq. \ref{heightvertex} becomes relevant in the algebraic
phase. The relevance of this vertex operator can be manifested by
directly calculating the partition function with expansion of the
vertex operator. The second order perturbation involves the
integral of the correlation function Eq. \ref{heightcorrelation},
and if $K > K_{c1}$, the integral diverges in the infrared limit.

the proliferation of the vertex operator will drive a
Kosterlitz-Thouless (KT) transition, across which the algebraic
correlation disappears, and the correlation length of the vertex
operators diverges as $\xi \sim \exp[\sqrt{\frac{c}{K -
K_{c1}}}]$. This KT transition in 3d space is due to the special
quasilocal symmetry in Eq. \ref{quasilocal}, which grants the 3d
system a 2d like symmetry for each flavor of $h_\mu$. The KT
transition and dimensional reduction behavior in 3d or 2+1d were
also discussed in another type of U(1) rotor systems with similar
quasilocal symmetries \cite{bosemetal,rotonliquid,xudimension}.
What is different here is that, this dimensional reduction
behavior in our system is inherited from the generic quasi-local
symmetry in the definition of height field Eq. \ref{height},
therefore dimensional reduction is robust.

There is another useful duality of Eq. \ref{heightfield}. Consider
the following Gaussian field theory: \beqn \mathcal{F} = \sum_{a =
1}^N \frac{1}{2K}[(\nabla_x\nabla_y \theta^a)^2 +
(\nabla_y\nabla_z \theta^a)^2 + (\nabla_z\nabla_x \theta^a)^2],
\label{rotorfield}\eeqn Now let us introduce the new field
$\phi^a_{\mu\nu}$ through Hubbard-Stratonovich transformation:
\beqn \mathcal{F} = \sum_{\mu\neq \nu, a}
\frac{K}{2}(\phi^a_{\mu\nu})^2 -
i\phi^a_{\mu\nu}\nabla_{\mu}\nabla_{\nu}\theta^a. \eeqn After
integrating out the field $\theta^a$, the partition function of
the system becomes \beqn Z = \int
D\phi^a_{\mu\nu}\delta(\sum_{\mu\neq\nu}\nabla_\mu\nabla_\nu
\phi^a_{\mu\nu})e^{- \int d^3x \sum_{a}
\sum_{\mu\neq\nu}\frac{K}{2}(\phi^a_{\mu\nu})^2}.
\label{onrotor}\eeqn The delta function in this partition imposes
the same constraint Eq. \ref{phi2constraint}. The field theory Eq.
\ref{rotorfield} is also invariant under quasi-local symmetry
transformation: \beqn \theta^a \rightarrow \theta^a + g^a_1(x) +
g^a_2(y) + g^a_3(z). \label{quasilocal2}\eeqn $g^a_\alpha$ are
arbitrary functions of one of the three coordinates. The field
theories Eq. \ref{rotorfield} and Eq. \ref{heightfield} are dual
to each other, with the same parameter $K$.

\section{IV, Defects of Constraints}

\subsection{A, Affects on the Algebraic Phase}

So far all the constraints have been perfectly imposed, and in
this section we will consider the case with slightly softened
constraint \cite{sondhimessage} $i.e.$ we allow the existence of
the point defect that violates the constraint. In the algebraic
phase, for all the O($N$) models considered in this work, if $g$
is large but finite in Eq. \ref{field}, \ref{field2},
\ref{field3}, the power-law correlation immediately crossovers
back to short range correlation for large enough distance. For
instance, take the correlation function Eq. \ref{correlationg}, we
can see that the length scale for this crossover is $l \sim
\sqrt{g/r}$ for O($N$) spin-ice. 

The Gaussian theory evaluation of the defects is based on the
assumption that the defect can take continuous values, therefore
the ``charge" of the defect $\sum_\mu \nabla_\mu\phi^a_\mu$ can be
infinitesimal. Therefore the Gaussian theory is no longer
applicable for $N = 1$, where the defect charge is always
discrete. To evaluate the defect in this case, one needs to go to
the other side of the duality. For instance, the CDM-$m$ is dual
to the Villain form of U(1) rotor model with partition function
\beqn Z = \int D\theta \sum_{l_{i,\mu}}\exp[ &-& \sum_{i,\mu}
\frac{1}{2\tilde{K}}(\theta_{i} - \theta_{i + \mu}- 2\pi
l_{i,\mu})^2 \cr\cr &+& 2\pi i \frac{m}{6}l_{i,\mu}(-1)^i]. \eeqn
Notice that there is an imaginary term in the partition function
due to the nonzero average filling of the dimer density, which is
similar to the Berry phase of quantum Bose rotor model with
fractional boson filling \cite{powell2}. To show this duality
explicitly, we still introduce the field $\phi_{i,\mu}$ through
Hubbard-Stratonovich transformation \beqn Z &=& \int D\theta
D\phi_{i,\mu} \sum_{l_{i,\mu}}\exp[ \sum_{i,\mu} -
\frac{\tilde{K}}{2}\phi^2_{i,\mu} - i\nabla_\mu
\phi_{i,\mu}\theta_i \cr\cr &-& 2\pi i l_{i,\mu}(\phi_{i,\mu} +
(-1)^i\frac{m}{6})]. \eeqn After integrating out the $\theta_i$,
and summing over $l_{i,\mu}$, the dual theory reads \beqn Z =
\sum_{\phi_{\mu}}\delta(\sum_{\mu}\nabla_\mu \phi_{\mu})\exp[
\sum_{i,\mu} - \frac{\tilde{K}}{2}(\phi_{i,\mu})^2], \eeqn and
$(-1)^i\phi_{i,\mu} \in \mathbb{Z} + m/6 $. Then any low energy
configuration of $\phi_{i,\mu}$ is equivalent to a CDM-$m$
configuration.

If we are in the superfluid phase of the rotor model, which is
dual to the algebraic phase of the CDM-$m$, we can expand the free
energy of the rotor model Eq. \ref{rotorfield} at $l_{i,\mu} = 0$,
and the field theory of the superfluid phase is simply
$\mathcal{F} = \frac{1}{2K}(\nabla_\mu\theta)^2$. The defects will
be taken into account by the vertex operator in the field theory:
\beqn \mathcal{F} = \sum_{\mu} \frac{1}{2K}(\nabla_\mu \theta)^2 -
\alpha \cos(\theta), \eeqn $\alpha \sim \exp(- \mathcal{C}_1 g)$
is the fugacity of the defect, and the partition function with
expansion of $\alpha$ is equivalent to a classical Coulomb gas,
which is equivalent to the partition function of defects. In 3d
space since $\theta$ has long range order, $\alpha$ is a very
relevant perturbation $i.e.$ the presence of defects will
immediately destroy the algebraic phase with infinitesimal
fugacity. Another way to show the relevance of the defect is to
compare the energy and entropy of an isolated defect. If the
system size is $L^3$, then an isolated defect in the algebraic
phase costs energy $E \sim \int d^3k \frac{1}{k^2} \sim
g\mathcal{A} + \frac{1}{L}$ which is finite in the infrared limit.
While the entropy of the defect scales as $S \sim \ln L$,
therefore the entropy always dominates energy $i.e.$ the defects
always proliferate.

The situation is very different for the CPM-$m$. Just like the
duality of the CDM-$m$, the CPM-$m$ is dual to the following rotor
model: \beqn Z = \int D\theta \sum_{l_{i,\mu\nu}}\exp[ &-&
\sum_{i,\mu\neq\nu} \frac{1}{2\tilde{K}}(\nabla_{\mu\nu}\theta -
2\pi l_{i,\mu\nu})^2 \cr\cr &+& 2\pi i \frac{m}{12}
l_{i,\mu\nu}(-1)^i]. \eeqn Again, in the algebraic phase of the
CPM-$m$, we can expand the free energy at $l_{i,\mu\nu} = 0$, and
evaluate the relevance of defects in the rotor field theory: \beqn
\mathcal{F} = \sum_{\mu\neq \nu}
\frac{1}{2K}(\nabla_\mu\nabla_\nu\theta)^2 - \alpha \cos(\theta).
\label{rotorfield2}\eeqn Now because of the quasi-local symmetry
Eq. \ref{quasilocal2}, the vertex operator $\mathcal{F}_v = -
\alpha\cos(\theta)$ has no nonzero correlation spatially, which
seemingly implies that the defect is irrelevant. This effect can
again be shown by evaluating the energy and entropy of the defect.
An isolated defect costs energy \beqn E \sim \int d^3k
\frac{1}{k_x^2k_y^2 + k_y^2k_z^2 + k_z^2k_x^2} \sim L + c \ln L,
\eeqn which always dominates the entropy that scales as $\ln L$
$i.e.$ the defects are always suppressed by energy.

However, although nonzero correlation between $\mathcal{F}_v$ is
forbidden by symmetry, the correlation between defect-dipole
operator $\mathcal{F}_{d,\mu} \sim -\alpha^\prime \cos(\nabla_\mu
\theta) = -\alpha^\prime \cos(\theta_i - \theta_{i+\mu})$ can be
nonzero. For instance, $\mathcal{F}_{d,z}$ can have nonzero
correlation within the entire $xy$ plane: \beqn && \langle
\mathcal{F}_{d,z}(0,0,0)\mathcal{F}_{d,z}(x,y,0) \rangle \cr\cr
&\sim& \exp\{ - K \int \frac{d^3k}{(2\pi)^3} \frac{k_z^2e^{ik_x x
+ik_y y}}{k_x^2k_y^2 + k_y^2k_z^2 + k_x^2k_z^2} \} \cr\cr &\sim&
\frac{1}{(x^2+y^2)^{K/(4\pi)}}. \eeqn When $K < K_{c2} = 8\pi$,
the vertex dipole operator $\mathcal{F}_{d, \mu}$ is relevant, and
the algebraic phase disappears. Suppose $\alpha^\prime \sim
\alpha^2 \sim \exp(- \mathcal{C}_2 g)$ is the fugacity of the
defect dipole, the crossover length scale beyond which the dipole
becomes important is $\xi \sim (\alpha^\prime)^{4\pi/(K_{c2} -
K)}$. Again we can compare the energy of a defect dipole and its
entropy. The energy of a defect dipole scales as $ E \sim \int
d^3k k_z^2/(k_x^2k_y^2 + k_y^2k_z^2 + k_z^2k_x^2) \sim \ln L $,
which is comparable with the entropy. So by tuning $K$ in Eq.
\ref{rotorfield2}, there will be a KT transition as a result of
the competition between entropy and energy.

\subsection{B, Effects on the Transition}

Close to the transition $r = 0$, we can use the correlation
function Eq. \ref{correlationg} to compute the RG equation. $g$
may flow under RG eventually, but it remains a constant at the
first order $\epsilon$ expansion, and hence we will just take it a
constant. For the isotropic O($N$) spin-ice, the RG equation takes
the same form as Eq. \ref{staggeredRG}, but now we need to
reevaluate the loop integrals: \beqn A &=& \frac{5}{8} +
\frac{1}{8(1+g)^2} + \frac{1}{4(1+g)}, \cr\cr B &=&
\frac{g^2}{24(1+g)^2}, \cr\cr C &=& \frac{13}{24} + \frac{1}{24(1
+ g)^2} + \frac{5}{12(1 + g)}. \label{abcg}\eeqn Solve the
equation Eq. \ref{staggeredRG} with the new parameters Eq.
\ref{abcg}, we can see that when $g$ is large, the solutions are
qualitatively unchanged from Eq. \ref{fixedpoint2}. While when $g
< g_c \sim 4.1$, stable fixed points are found with $N = 1$, which
corresponds to CDM-3 with softened constraint. For instance, with
$N = 1$, $g = 1$, the stable fixed point is located at $u_\ast =
\frac{53-\sqrt{37}}{3330}$, $v_\ast + w_\ast =
\frac{7+\sqrt{37}}{555}$. Now in the low temperature phase the
connected correlation on top of the long range order is also
short-ranged after the transition described above.

With finite $g$, the O($N$) spin-ice with easy-plane axis can be
studied in a similar way as last paragraph. As was mentioned in
section II. A, we need to generalize the system to 4d with $\mu =
x, \ y, \ z, \ \tau$, and anisotropy of the Hamiltonian prefers
the O($N$) spin vectors on $x, \ y, \ z$ axes to order first at
$T_{c1}$. At the critical point $T_{c1}$ with $r = 0$, the 4d
correlation function between $\phi^a_\mu$ reads \beqn  \lim_{m
\rightarrow \infty, r \rightarrow 0} D^{ab}_{\mu\nu} &\sim&
\frac{\delta_{ab}}{q^2}(\delta_{\mu\nu} - \frac{g q_\mu
q_\nu}{q_\tau^2 + (1 + g)\mathbf{q}^2}) \ , \mu, \nu \neq \tau,
\cr\cr \mathrm{\mathbf{q}}^2 &=& q_x^2 + q_y^2 + q_\tau^2, \cr\cr
\lim_{m \rightarrow \infty, r \rightarrow 0} D^{ab}_{\mu\nu} &=&
0, \ \mu \ \mathrm{or} \ \nu = 0. \eeqn The RG equation in this
case takes the same form as Eq. \ref{staggeredRG2}, with different
$A$, $B$ and $C$. Expanded to the first order of $1/g$, these
parameters read: \beqn A &=& \frac{8}{15} + \frac{8}{15g} +
O(\frac{1}{g^2}), \cr\cr B &=& \frac{1}{15} - \frac{4}{15g} +
O(\frac{1}{g^2}), \cr\cr C &=& \frac{2}{5} + \frac{16}{15g} +
O(\frac{1}{g^2}). \eeqn When $g < g_c \sim 4.4$, stable fixed
points are found with $N = 1$, which corresponds to CDM-3 with
softened constraint and easy plane anisotropy. With easy-axis
anisotropy and finite $g$, the effective $z = 2$ physics in Eq.
\ref{z22} is absent, and the universality class is expected to
crossover back to a 3d Wilson-Fisher transition.

In the O($N$) spin-plaquette model, since the $g$ term has higher
derivatives compared with the Gaussian part of the field theory
Eq. \ref{field3}, then $g$ becomes irrelevant when its initial
value is finite. Since at the low energy effective field theory,
there is no flavor-orbital mixing interaction, we can take a more
concise notation: $\phi^a_{x} = \phi^a_{yz}$, $\phi^a_{y} =
\phi^a_{zx}$, $\phi^a_{z} = \phi^a_{xy}$. At the critical point $r
= 0$ the O($N$) spin-plaquette model is described by the field
theory: \beqn \mathcal{F} = \sum_{\mu, a} - \phi^a_{\mu} \nabla^2
\phi^a_{\mu} + \mathcal{F}_4, \label{field4}\eeqn with
$\mathcal{F}_4$ given by Eq. \ref{f4}. Now the coupled RG equation
for $u$, $v$ and $w$ reads \beqn \frac{d u }{d\ln l} &=& \epsilon
u - 8(8 + N)u^2 - 4Nv^2 - 4w^2 - 8vw, \cr\cr \frac{d v }{d\ln l}
&=& \epsilon v - 4(4 + N) v^2 - 4w^2 \cr\cr &-& 16(2 + N)uv - 24
vw - 16 uw, \cr\cr \frac{d w }{d\ln l} &=& \epsilon w - 4(3 + N)
w^2- 16 vw - 32 uw, \cr\cr \frac{dr}{d\ln l} &=& 2 r - 8(2+N)ur -
8Nvr - 8wr. \label{columnarRG} \eeqn For large enough $N$, there
are in total eight fixed points. For instance, the fixed point
$w_\ast = 0$, $v_\ast = 2 u_\ast = \epsilon/(12N) + \cdots$ has
enlarged O(3$N$) symmetry, which mixes the flavor and spin
symmetry. Expanded to the order of $\epsilon/N^2$, the only stable
fixed point with $r = 0$ is located at \beqn u_\ast &=&
\frac{\epsilon}{8N} -\frac{5\epsilon}{4N^2} +
O(\frac{\epsilon}{N^2}), \cr\cr \ v_\ast &=& - \frac{3\epsilon}{4
N^2} + O(\frac{\epsilon}{N^2}) , \cr\cr \ w_\ast &=&
\frac{\epsilon}{4N} - \frac{7\epsilon}{4 N^2} +
O(\frac{\epsilon}{N^2}). \eeqn Close to the stable fixed points,
the three eigenvectors of the RG flow have scaling dimensions
\beqn \Delta_1 &=& - \epsilon + O(\frac{\epsilon}{N^2}), \cr\cr
\Delta_2 &=& - \epsilon + \frac{4\epsilon}{N} +
O(\frac{\epsilon}{N^2}), \cr\cr \Delta_3 &=& - \epsilon +
\frac{20\epsilon}{N} + O(\frac{\epsilon}{N^2}). \eeqn The scaling
dimension of $r$ at this fixed point is \beqn [r] = \frac{1}{\nu}
= 2 - \epsilon + \frac{12\epsilon}{N} + O(\frac{\epsilon}{N^2},
\epsilon^2). \eeqn

If we take $N = 1$, the field theory Eq. \ref{field4} becomes the
standard O(3) transition with cubic anisotropy, and various
numerical methods have confirmed that the 3d O(3) symmetric fixed
point is stable \cite{vicari2003}, with critical exponent $[r] =
1/\nu = 2 - 5\epsilon/11 + O(\epsilon^2)$ (see Ref. \cite{amit}).

\section{V, Summaries and Discussions}

In this work we studied the classical phase transition between the
algebraic phase and low temperature spin ordered phase in several
different types of O($N$) spin models with local geometric
constraint. Effects of softened constraints are also considered in
all the models. Systematic RG calculations are applied to all of
the cases, and solutions at the first order $\epsilon$ expansion
were obtained as precisely as we could. However, higher order
$\epsilon$ expansions, as well as direct Monte Carlo simulations
of the lattice models are indeed demanded in order to confirm our
results at a more quantitative level.

So far we have been focusing on the staggered spin order in all
cases we studied, which has a straightforward order parameter
description. However, there are another large class of spin orders
of constrained systems that seem to involve more complicated order
parameter descriptions. For instance, in the columnar order of
spin-ice, $\phi^a_\mu$ does not have nonzero expectation values,
therefore we need to develop another formalism for this case. The
columnar order is equivalent to order of $\phi^a_x$ at momentum
$Q_1 = (0, \pi, \pi)$, $\phi^a_y$ at momentum $Q_1 = (\pi, 0,
\pi)$ and $\phi^a_z$ at momentum $Q_1 = (\pi, \pi, 0)$. Therefore
presumably we could describe this transition with condensation of
$\phi^a_\mu$ at all three wave-vectors. However, just like the
Coulomb-columnar transition in CDM-1 discussed in section II-A,
some topological configuration of these order parameters may be
forbidden, which potentially can change the universality class
completely.

Another interesting subject is to generalize our formalism to the
quantum case. For instance, it is well-known that the quantum
O($N$) rotor model and spin models can be described by nonlinear
sigma model in the infrared limit. Suppose we impose a local
constraint on the order parameters of the quantum rotor model or
spin model, it is possible that we can use field theories similar
to Eq. \ref{field}, \ref{field2} to describe the quantum phase
transitions in the constrained Nonlinear sigma model: \beqn
\mathcal{L} &=& \sum_{a}\sum_{\mu = x, y \cdots
}\frac{1}{2g}[(\partial_\tau n^a_\mu)^2 + v^2 (\vec{\nabla}
n^a_\mu)^2] \cr\cr &+& \sum_a \beta(\sum_{\mu = x, y, \cdots}
\nabla_\mu n^a_\mu)^2 + \cdots \eeqn with the limit $\beta
\rightarrow \infty$. Notice that flavor $\mu$ takes only spatial
coordinates.


\bibliography{dimer}

\begin{thebibliography}{39}
\expandafter\ifx\csname natexlab\endcsname\relax\def\natexlab#1{#1}\fi
\expandafter\ifx\csname bibnamefont\endcsname\relax
  \def\bibnamefont#1{#1}\fi
\expandafter\ifx\csname bibfnamefont\endcsname\relax
  \def\bibfnamefont#1{#1}\fi
\expandafter\ifx\csname citenamefont\endcsname\relax
  \def\citenamefont#1{#1}\fi
\expandafter\ifx\csname url\endcsname\relax
  \def\url#1{\texttt{#1}}\fi
\expandafter\ifx\csname urlprefix\endcsname\relax\def\urlprefix{URL }\fi
\providecommand{\bibinfo}[2]{#2}
\providecommand{\eprint}[2][]{\url{#2}}

\bibitem[{\citenamefont{Moessner and
  Sondhi}(2003{\natexlab{a}})}]{sondhiphoton}
\bibinfo{author}{\bibfnamefont{R.}~\bibnamefont{Moessner}} \bibnamefont{and}
  \bibinfo{author}{\bibfnamefont{S.~L.} \bibnamefont{Sondhi}},
  \bibinfo{journal}{Phys. Rev. B} \textbf{\bibinfo{volume}{68}},
  \bibinfo{pages}{184512} (\bibinfo{year}{2003}{\natexlab{a}}).

\bibitem[{\citenamefont{Hermele et~al.}(2004)\citenamefont{Hermele, Fisher, and
  Balents}}]{hermelephoton}
\bibinfo{author}{\bibfnamefont{M.}~\bibnamefont{Hermele}},
  \bibinfo{author}{\bibfnamefont{M.~P.} \bibnamefont{Fisher}},
  \bibnamefont{and} \bibinfo{author}{\bibfnamefont{L.}~\bibnamefont{Balents}},
  \bibinfo{journal}{Phys. Rev. B} \textbf{\bibinfo{volume}{69}},
  \bibinfo{pages}{064404} (\bibinfo{year}{2004}).

\bibitem[{\citenamefont{Wen}(2003)}]{wenphoton}
\bibinfo{author}{\bibfnamefont{X.-G.} \bibnamefont{Wen}},
  \bibinfo{journal}{Phys. Rev. B} \textbf{\bibinfo{volume}{68}},
  \bibinfo{pages}{115413} (\bibinfo{year}{2003}).

\bibitem[{\citenamefont{Xu}(2006{\natexlab{a}})}]{xugraviton1}
\bibinfo{author}{\bibfnamefont{C.}~\bibnamefont{Xu}},
  \bibinfo{journal}{cond-mat/0602443}  (\bibinfo{year}{2006}{\natexlab{a}}).

\bibitem[{\citenamefont{Xu}(2006{\natexlab{b}})}]{xugraviton2}
\bibinfo{author}{\bibfnamefont{C.}~\bibnamefont{Xu}}, \bibinfo{journal}{Phys.
  Rev. B} \textbf{\bibinfo{volume}{74}}, \bibinfo{pages}{224433}
  (\bibinfo{year}{2006}{\natexlab{b}}).

\bibitem[{\citenamefont{Isakov et~al.}(2005)\citenamefont{Isakov, Moessner, and
  Sondhi}}]{sondhidipole}
\bibinfo{author}{\bibfnamefont{S.~V.} \bibnamefont{Isakov}},
  \bibinfo{author}{\bibfnamefont{R.}~\bibnamefont{Moessner}}, \bibnamefont{and}
  \bibinfo{author}{\bibfnamefont{S.~L.} \bibnamefont{Sondhi}},
  \bibinfo{journal}{Phys. Rev. Lett.} \textbf{\bibinfo{volume}{95}},
  \bibinfo{pages}{217201} (\bibinfo{year}{2005}).

\bibitem[{\citenamefont{Castelnovo et~al.}(2008)\citenamefont{Castelnovo,
  Moessner, and Sondhi}}]{sondhispinice2}
\bibinfo{author}{\bibfnamefont{C.}~\bibnamefont{Castelnovo}},
  \bibinfo{author}{\bibfnamefont{R.}~\bibnamefont{Moessner}}, \bibnamefont{and}
  \bibinfo{author}{\bibfnamefont{S.~L.} \bibnamefont{Sondhi}},
  \bibinfo{journal}{Nature} \textbf{\bibinfo{volume}{451}}, \bibinfo{pages}{42}
  (\bibinfo{year}{2008}).

\bibitem[{\citenamefont{Huse et~al.}(2003)\citenamefont{Huse, Krauth, Moessner,
  and Sondhi}}]{sondhidimer}
\bibinfo{author}{\bibfnamefont{D.~A.} \bibnamefont{Huse}},
  \bibinfo{author}{\bibfnamefont{W.}~\bibnamefont{Krauth}},
  \bibinfo{author}{\bibfnamefont{R.}~\bibnamefont{Moessner}}, \bibnamefont{and}
  \bibinfo{author}{\bibfnamefont{S.~L.} \bibnamefont{Sondhi}},
  \bibinfo{journal}{Phys. Rev. Lett.} \textbf{\bibinfo{volume}{91}},
  \bibinfo{pages}{167004} (\bibinfo{year}{2003}).

\bibitem[{\citenamefont{Chen et~al.}(2009)\citenamefont{Chen, Gukelberger,
  Trebst, Alet, and Balents}}]{balentsdimer}
\bibinfo{author}{\bibfnamefont{G.}~\bibnamefont{Chen}},
  \bibinfo{author}{\bibfnamefont{J.}~\bibnamefont{Gukelberger}},
  \bibinfo{author}{\bibfnamefont{S.}~\bibnamefont{Trebst}},
  \bibinfo{author}{\bibfnamefont{F.}~\bibnamefont{Alet}}, \bibnamefont{and}
  \bibinfo{author}{\bibfnamefont{L.}~\bibnamefont{Balents}},
  \bibinfo{journal}{Phys. Rev. B} \textbf{\bibinfo{volume}{80}},
  \bibinfo{pages}{045112} (\bibinfo{year}{2009}).

\bibitem[{\citenamefont{Alet et~al.}(2006)\citenamefont{Alet, Misguich,
  Pasquier, Moessner, and Jacobsen}}]{dimersimulation1}
\bibinfo{author}{\bibfnamefont{F.}~\bibnamefont{Alet}},
  \bibinfo{author}{\bibfnamefont{G.}~\bibnamefont{Misguich}},
  \bibinfo{author}{\bibfnamefont{V.}~\bibnamefont{Pasquier}},
  \bibinfo{author}{\bibfnamefont{R.}~\bibnamefont{Moessner}}, \bibnamefont{and}
  \bibinfo{author}{\bibfnamefont{J.~L.} \bibnamefont{Jacobsen}},
  \bibinfo{journal}{Phys. Rev. Lett.} \textbf{\bibinfo{volume}{97}},
  \bibinfo{pages}{030403} (\bibinfo{year}{2006}).

\bibitem[{\citenamefont{Misguich et~al.}(2008)\citenamefont{Misguich, Pasquier,
  and Alet}}]{dimersimulation2}
\bibinfo{author}{\bibfnamefont{G.}~\bibnamefont{Misguich}},
  \bibinfo{author}{\bibfnamefont{V.}~\bibnamefont{Pasquier}}, \bibnamefont{and}
  \bibinfo{author}{\bibfnamefont{F.}~\bibnamefont{Alet}},
  \bibinfo{journal}{Phys. Rev. B} \textbf{\bibinfo{volume}{78}},
  \bibinfo{pages}{100402(R)} (\bibinfo{year}{2008}).

\bibitem[{\citenamefont{Pauling}(1935)}]{pauling}
\bibinfo{author}{\bibfnamefont{L.}~\bibnamefont{Pauling}},
  \bibinfo{journal}{Journal of the American Chemical Society}
  \textbf{\bibinfo{volume}{57}}, \bibinfo{pages}{2680} (\bibinfo{year}{1935}).

\bibitem[{\citenamefont{Fennell et~al.}(2009)\citenamefont{Fennell, Deen,
  Wildes, Schmalzl, Prabhakaran, Boothroyd, Aldus, McMorrow, and
  Bramwell}}]{spinicemonopole1}
\bibinfo{author}{\bibfnamefont{T.}~\bibnamefont{Fennell}},
  \bibinfo{author}{\bibfnamefont{P.~P.} \bibnamefont{Deen}},
  \bibinfo{author}{\bibfnamefont{A.~R.} \bibnamefont{Wildes}},
  \bibinfo{author}{\bibfnamefont{K.}~\bibnamefont{Schmalzl}},
  \bibinfo{author}{\bibfnamefont{D.}~\bibnamefont{Prabhakaran}},
  \bibinfo{author}{\bibfnamefont{A.~T.} \bibnamefont{Boothroyd}},
  \bibinfo{author}{\bibfnamefont{R.~J.} \bibnamefont{Aldus}},
  \bibinfo{author}{\bibfnamefont{D.~F.} \bibnamefont{McMorrow}},
  \bibnamefont{and} \bibinfo{author}{\bibfnamefont{S.~T.}
  \bibnamefont{Bramwell}}, \bibinfo{journal}{Science}
  \textbf{\bibinfo{volume}{326}}, \bibinfo{pages}{415} (\bibinfo{year}{2009}).

\bibitem[{\citenamefont{Moessner and Sondhi}(2003{\natexlab{b}})}]{sondhi2003}
\bibinfo{author}{\bibfnamefont{R.}~\bibnamefont{Moessner}} \bibnamefont{and}
  \bibinfo{author}{\bibfnamefont{S.~L.} \bibnamefont{Sondhi}},
  \bibinfo{journal}{Phys. Rev. B} \textbf{\bibinfo{volume}{68}},
  \bibinfo{pages}{184512} (\bibinfo{year}{2003}{\natexlab{b}}).

\bibitem[{\citenamefont{Hermele et~al.}(2005)\citenamefont{Hermele, Senthil,
  and M.P.A.Fisher}}]{Hermele2005}
\bibinfo{author}{\bibfnamefont{M.}~\bibnamefont{Hermele}},
  \bibinfo{author}{\bibfnamefont{T.}~\bibnamefont{Senthil}}, \bibnamefont{and}
  \bibinfo{author}{\bibnamefont{M.P.A.Fisher}}, \bibinfo{journal}{Phys. Rev.
  B.} \textbf{\bibinfo{volume}{72}}, \bibinfo{pages}{104404}
  (\bibinfo{year}{2005}).

\bibitem[{\citenamefont{Pankov et~al.}(2007)\citenamefont{Pankov, Moessner, and
  Sondhi}}]{pankov2007}
\bibinfo{author}{\bibfnamefont{S.}~\bibnamefont{Pankov}},
  \bibinfo{author}{\bibfnamefont{R.}~\bibnamefont{Moessner}}, \bibnamefont{and}
  \bibinfo{author}{\bibfnamefont{S.~L.} \bibnamefont{Sondhi}},
  \bibinfo{journal}{Phys. Rev. B} \textbf{\bibinfo{volume}{76}},
  \bibinfo{pages}{104436} (\bibinfo{year}{2007}).

\bibitem[{\citenamefont{Xu and Wu}(2008)}]{xuwu2008}
\bibinfo{author}{\bibfnamefont{C.}~\bibnamefont{Xu}} \bibnamefont{and}
  \bibinfo{author}{\bibfnamefont{C.}~\bibnamefont{Wu}}, \bibinfo{journal}{Phys.
  Rev. B} \textbf{\bibinfo{volume}{77}}, \bibinfo{pages}{134449}
  (\bibinfo{year}{2008}).

\bibitem[{\citenamefont{Wu et~al.}(2003)\citenamefont{Wu, Hu, and
  Zhang}}]{wu2003}
\bibinfo{author}{\bibfnamefont{C.}~\bibnamefont{Wu}},
  \bibinfo{author}{\bibfnamefont{J.~P.} \bibnamefont{Hu}}, \bibnamefont{and}
  \bibinfo{author}{\bibfnamefont{S.~C.} \bibnamefont{Zhang}},
  \bibinfo{journal}{Phys. Rev. Lett} \textbf{\bibinfo{volume}{91}},
  \bibinfo{pages}{186402} (\bibinfo{year}{2003}).

\bibitem[{\citenamefont{Gorshkov et~al.}(2009)\citenamefont{Gorshkov, Hermele,
  Gurarie, Xu, Julienne, Ye, Zoller, Demler, Lukin, and Rey}}]{gorshkov2009}
\bibinfo{author}{\bibfnamefont{A.~V.} \bibnamefont{Gorshkov}},
  \bibinfo{author}{\bibfnamefont{M.}~\bibnamefont{Hermele}},
  \bibinfo{author}{\bibfnamefont{V.}~\bibnamefont{Gurarie}},
  \bibinfo{author}{\bibfnamefont{C.}~\bibnamefont{Xu}},
  \bibinfo{author}{\bibfnamefont{P.~S.} \bibnamefont{Julienne}},
  \bibinfo{author}{\bibfnamefont{J.}~\bibnamefont{Ye}},
  \bibinfo{author}{\bibfnamefont{P.}~\bibnamefont{Zoller}},
  \bibinfo{author}{\bibfnamefont{E.}~\bibnamefont{Demler}},
  \bibinfo{author}{\bibfnamefont{M.~D.} \bibnamefont{Lukin}}, \bibnamefont{and}
  \bibinfo{author}{\bibfnamefont{A.~M.} \bibnamefont{Rey}},
  \bibinfo{journal}{arXiv:0905.2610}  (\bibinfo{year}{2009}).

\bibitem[{\citenamefont{Motrunich and Senthil}(2005)}]{senthilmotrunich}
\bibinfo{author}{\bibfnamefont{O.~I.} \bibnamefont{Motrunich}}
  \bibnamefont{and} \bibinfo{author}{\bibfnamefont{T.}~\bibnamefont{Senthil}},
  \bibinfo{journal}{Phys. Rev. B} \textbf{\bibinfo{volume}{71}},
  \bibinfo{pages}{125102} (\bibinfo{year}{2005}).

\bibitem[{\citenamefont{Powell and Chalker}(2008)}]{powell1}
\bibinfo{author}{\bibfnamefont{S.}~\bibnamefont{Powell}} \bibnamefont{and}
  \bibinfo{author}{\bibfnamefont{J.~T.} \bibnamefont{Chalker}},
  \bibinfo{journal}{Phys. Rev. Lett.} \textbf{\bibinfo{volume}{101}},
  \bibinfo{pages}{155702} (\bibinfo{year}{2008}).

\bibitem[{\citenamefont{Motrunich and Vishwanath}(2004)}]{motrunichashvin}
\bibinfo{author}{\bibfnamefont{O.~I.} \bibnamefont{Motrunich}}
  \bibnamefont{and}
  \bibinfo{author}{\bibfnamefont{A.}~\bibnamefont{Vishwanath}},
  \bibinfo{journal}{Phys. Rev. B} \textbf{\bibinfo{volume}{70}},
  \bibinfo{pages}{075104} (\bibinfo{year}{2004}).

\bibitem[{\citenamefont{Kamal and Murthy}(1993)}]{murthy1993}
\bibinfo{author}{\bibfnamefont{M.}~\bibnamefont{Kamal}} \bibnamefont{and}
  \bibinfo{author}{\bibfnamefont{G.}~\bibnamefont{Murthy}},
  \bibinfo{journal}{Phys. Rev. Lett.} \textbf{\bibinfo{volume}{71}},
  \bibinfo{pages}{1911} (\bibinfo{year}{1993}).

\bibitem[{\citenamefont{Motrunich and Vishwanath}(2009)}]{motrunichashvin2}
\bibinfo{author}{\bibfnamefont{O.~I.} \bibnamefont{Motrunich}}
  \bibnamefont{and}
  \bibinfo{author}{\bibfnamefont{A.}~\bibnamefont{Vishwanath}},
  \bibinfo{journal}{arXiv:0805.1494}  (\bibinfo{year}{2009}).

\bibitem[{\citenamefont{Henley}(1989)}]{henley1989}
\bibinfo{author}{\bibfnamefont{C.~L.} \bibnamefont{Henley}},
  \bibinfo{journal}{Phys. Rev. Lett.} \textbf{\bibinfo{volume}{62}},
  \bibinfo{pages}{2056} (\bibinfo{year}{1989}).

\bibitem[{\citenamefont{Aharony}(1975)}]{aharony1975}
\bibinfo{author}{\bibfnamefont{A.}~\bibnamefont{Aharony}},
  \bibinfo{journal}{Phys. Rev. B} \textbf{\bibinfo{volume}{12}},
  \bibinfo{pages}{1049} (\bibinfo{year}{1975}).

\bibitem[{bal()}]{balentsnote}
\bibinfo{howpublished}{The author is grateful to Leon Balents for pointing this
  out.}

\bibitem[{\citenamefont{Fisher and Aharony}(1973)}]{dipolar1}
\bibinfo{author}{\bibfnamefont{M.~E.} \bibnamefont{Fisher}} \bibnamefont{and}
  \bibinfo{author}{\bibfnamefont{A.}~\bibnamefont{Aharony}},
  \bibinfo{journal}{Phys. Rev. Lett.} \textbf{\bibinfo{volume}{30}},
  \bibinfo{pages}{559} (\bibinfo{year}{1973}).

\bibitem[{\citenamefont{Aharony and Fisher}(1973)}]{dipolar2}
\bibinfo{author}{\bibfnamefont{A.}~\bibnamefont{Aharony}} \bibnamefont{and}
  \bibinfo{author}{\bibfnamefont{M.~E.} \bibnamefont{Fisher}},
  \bibinfo{journal}{Phys. Rev. B} \textbf{\bibinfo{volume}{8}},
  \bibinfo{pages}{3323} (\bibinfo{year}{1973}).

\bibitem[{\citenamefont{Pickles et~al.}(2008)\citenamefont{Pickles, Saunders,
  and Chalker}}]{anisotropy}
\bibinfo{author}{\bibfnamefont{T.~S.} \bibnamefont{Pickles}},
  \bibinfo{author}{\bibfnamefont{T.~E.} \bibnamefont{Saunders}},
  \bibnamefont{and} \bibinfo{author}{\bibfnamefont{J.~T.}
  \bibnamefont{Chalker}}, \bibinfo{journal}{Europhysics Letters}
  \textbf{\bibinfo{volume}{84}}, \bibinfo{pages}{36002} (\bibinfo{year}{2008}).

\bibitem[{\citenamefont{Khaliullin and Maekawa}(2000)}]{khaliullin2000}
\bibinfo{author}{\bibfnamefont{G.}~\bibnamefont{Khaliullin}} \bibnamefont{and}
  \bibinfo{author}{\bibfnamefont{S.}~\bibnamefont{Maekawa}},
  \bibinfo{journal}{Phys. Rev. Lett} \textbf{\bibinfo{volume}{85}},
  \bibinfo{pages}{3950} (\bibinfo{year}{2000}).

\bibitem[{\citenamefont{Paramekanti et~al.}(2002)\citenamefont{Paramekanti,
  Balents, and Fisher}}]{bosemetal}
\bibinfo{author}{\bibfnamefont{A.}~\bibnamefont{Paramekanti}},
  \bibinfo{author}{\bibfnamefont{L.}~\bibnamefont{Balents}}, \bibnamefont{and}
  \bibinfo{author}{\bibfnamefont{M.~P.~A.} \bibnamefont{Fisher}},
  \bibinfo{journal}{Phys. Rev. B} \textbf{\bibinfo{volume}{66}},
  \bibinfo{pages}{054526} (\bibinfo{year}{2002}).

\bibitem[{\citenamefont{Balents and Fisher}(2005)}]{rotonliquid}
\bibinfo{author}{\bibfnamefont{L.}~\bibnamefont{Balents}} \bibnamefont{and}
  \bibinfo{author}{\bibfnamefont{M.~P.~A.} \bibnamefont{Fisher}},
  \bibinfo{journal}{Phys. Rev. B} \textbf{\bibinfo{volume}{71}},
  \bibinfo{pages}{085119} (\bibinfo{year}{2005}).

\bibitem[{\citenamefont{Xu and Moore}(2005)}]{xudimension}
\bibinfo{author}{\bibfnamefont{C.}~\bibnamefont{Xu}} \bibnamefont{and}
  \bibinfo{author}{\bibfnamefont{J.~E.} \bibnamefont{Moore}},
  \bibinfo{journal}{Nucl. Phys. B} \textbf{\bibinfo{volume}{716}},
  \bibinfo{pages}{487} (\bibinfo{year}{2005}).

\bibitem[{son()}]{sondhimessage}
\bibinfo{howpublished}{The author thanks Shivaji Sondhi for suggesting me study
  the case with softened constraints.}

\bibitem[{\citenamefont{Powell and Chalker}(2009)}]{powell2}
\bibinfo{author}{\bibfnamefont{S.}~\bibnamefont{Powell}} \bibnamefont{and}
  \bibinfo{author}{\bibfnamefont{J.~T.} \bibnamefont{Chalker}},
  \bibinfo{journal}{Phys. Rev. B} \textbf{\bibinfo{volume}{80}},
  \bibinfo{pages}{134413} (\bibinfo{year}{2009}).

\bibitem[{\citenamefont{Calabrese et~al.}(2003)\citenamefont{Calabrese,
  Pelissetto, and Vicari}}]{vicari2003}
\bibinfo{author}{\bibfnamefont{P.}~\bibnamefont{Calabrese}},
  \bibinfo{author}{\bibfnamefont{A.}~\bibnamefont{Pelissetto}},
  \bibnamefont{and} \bibinfo{author}{\bibfnamefont{E.}~\bibnamefont{Vicari}},
  \bibinfo{journal}{arXiv:cond-mat/0306273}  (\bibinfo{year}{2003}).

\bibitem[{\citenamefont{Amit and Martin-Mayor}(2005)}]{amit}
\bibinfo{author}{\bibfnamefont{D.~J.} \bibnamefont{Amit}} \bibnamefont{and}
  \bibinfo{author}{\bibfnamefont{V.}~\bibnamefont{Martin-Mayor}},
  \emph{\bibinfo{title}{Field Theory, the Renormalization Group and Gritical
  Phenomena}} (\bibinfo{publisher}{World Scientific Publishing Company},
  \bibinfo{year}{2005}).

\bibitem[{\citenamefont{Xu}(2009)}]{xufuture}
\bibinfo{author}{\bibfnamefont{C.}~\bibnamefont{Xu}}, \bibinfo{journal}{In
  progress}  (\bibinfo{year}{2009}).

\end{thebibliography}

\end{document}